\documentclass[
reprint,
superscriptaddress,
amsmath,amssymb,
aps, 
prl,
]{revtex4-2}

\usepackage{newtxtext,newtxmath}
\usepackage{graphicx}
\usepackage{dcolumn}
\usepackage{bm}
\usepackage[dvipsnames,svgnames]{xcolor}
\usepackage{xr-hyper} 
\usepackage{hyperref}
\hypersetup{
    pdfnewwindow=true,          
    colorlinks=true,            
    linkcolor=Black,         
    citecolor=Black,         
    filecolor=Black,         
    urlcolor=Black           
}

\usepackage{tikz}   
\usetikzlibrary{chains}
\usepackage{tikz-3dplot}

\usepackage{amsmath,amssymb}
\usepackage{stmaryrd}
\usepackage{latexsym}
\usepackage{CJK}
\usepackage{graphicx}
\usepackage{indentfirst}
\usepackage{listings}
\usepackage{booktabs}
\usepackage{stmaryrd}
\usepackage{multirow}
\usepackage{verbatim}
\usepackage{makecell}
\usepackage{lineno,hyperref}
\usepackage{longtable}
\usepackage{hyperref}

\newcommand{\txR}{\text{R}}
\newcommand{\rmd}{\mathrm{d}}

\begin{document}

\title{A Generalized Frank-Bilby Equation for Interfaces in Crystalline Materials}

\author{Dongsong Tao}
\affiliation{Department of Materials Science and Engineering, City University of Hong Kong, Hong Kong SAR, China}
\author{Luchan Zhang}
\affiliation{College of Mathematics and Statistics, Shenzhen University, Shenzhen, China}
\author{David J. Srolovitz}
\affiliation{Department of Mechanical Engineering, The University of Hong Kong, Hong Kong SAR, China}
\author{Yang Xiang}
\affiliation{Department of Mathematics, The Hong Kong University of Science and Technology,  Hong Kong SAR, China}
\author{Jian Han}
\email{jianhan@cityu.edu.hk}
\affiliation{Department of Materials Science and Engineering, City University of Hong Kong, Hong Kong SAR, China}


\begin{abstract}
The classical Frank-Bilby equation (FBE) is commonly used to predict the structure of interfaces in crystalline materials in terms of interfacial dislocation networks. 
However, in general, the line defects in interfaces are disconnections, possessing \emph{both} dislocation and step character, which are not captured by the classical FBE. 
As a result, the FBE cannot fully describe the structure of most interfaces of practical interest.
To address this issue, we derive a generalized Frank-Bilby equation (GFBE) that explicitly incorporates both dislocation and step components of interfacial defects.
We demonstrate its application to several representative interface systems.
\end{abstract}

\maketitle

Interfaces are two-dimensional defects that separate crystalline phases or grains of different orientations. 
Predicting interface structure is essential for understanding interface motion and a wide range of material properties~\cite{SuttonBalluffi1995,hirth1982theory}.  
Although interface structure can be fully described at the atomic scale, it is often insightful to represent it as an ordered atomic structure interrupted by arrays of line defects~\cite{SuttonBalluffi1995, han2017grain, Han2018Grain}. 

Traditionally, these interfacial line defects are treated as dislocations. 
The Frank-Bilby equation (FBE)~\cite{frank1950report,bilby1955report,bilby1955continuous} predicts the interfacial Burgers vector content which is geometrically necessary to accommodate the mismatch between adjoining crystals and eliminate the associated far-field stress~\cite{Vattre2013b}. 
However, the FBE describes a continuous Burgers vector density rather than the discrete interfacial dislocations as observed experimentally. 
The quantized FBE (qFBE)~\cite{SuttonBalluffi1995,Yang2009Quantization,Sangghaleh2018AIDA} addresses this by interpreting the Burgers vector content in terms of interfacial dislocation networks. 

However, interfacial line defects are more appropriately described as \emph{disconnections}~\cite{Hirth1994Dislocationstep,Hirth2013Interface}, which possess both dislocation and step character~\footnote{Here, ``disconnection'' is used in a general sense to include all translational-symmetry-derived interfacial line defects. Interfacial dislocations and pure steps correspond to disconnections with zero step height and zero Burgers vector, respectively.}.
Because the present qFBE formulations account only for dislocation character, they cannot fully describe the structure of most interfaces.
Several recent approaches have attempted to address this limitation. 
Gordon and Sills~\cite{gordon2024self} solved the qFBE by adjusting the interface inclination to achieve consistency with disconnection step density; however, their formulation is applicable to a restricted set of defect configurations (i.e., two sets of interfacial dislocations and one set of disconnections) and does not generalize to cases where misorientation also needs to be adjusted. 
The O-lattice (O-line) model~\cite{bollmann1970crystal,Zhang1993Olattice2,Qiu2013Crystallography,Zhang2021study} provides a geometric description of lattice matching and has been widely used to analyze interfacial structure, but does not explicitly incorporate step character of interfacial line defects and therefore cannot directly describe disconnection networks.

Here, we derive a generalized Frank-Bilby equation (GFBE) that describes interfaces in terms of quantized disconnection networks. 
By explicitly incorporating both Burgers vector and step height components, the GFBE provides a simple, physically transparent, and broadly applicable framework for predicting interfacial structure.
It defines the limits of what can be inferred from interfacial geometry alone, without invoking energetic or kinetic information. 

Consider a bicrystal containing an infinitely large, planar interface separating ``$+$'' and ``$-$'' crystals; this configuration is termed the natural state. 
The unit normal vector $\hat{\bm{n}}$ points from the ``$-$'' to the ``$+$'' crystal. 
We introduce a reference state in which the two crystals are brought structurally coherent along the interface by suitable deformation and/or rotation. 
Let $\mathbf{F}^{+/-}$ be the deformation gradient mapping the reference state of the ``$+$''/``$-$'' crystal to its natural state. 
A Burgers vector content is required to accommodate the mismatch resulted from the transformation along the interface, measured by the jump: $\llbracket \mathbf{F}^{-1} \rrbracket = (\mathbf{F}^+)^{-1} - (\mathbf{F}^-)^{-1}$. 
The Nye tensor~\cite{Nye1953Some}, describing the interfacial Burgers vector density, is 
$\bm{\upalpha}
= \llbracket \mathbf{F}^{-1} \rrbracket
(\hat{\bm{n}}\times) \delta(\bm{x}\cdot\hat{\bm{n}} - c)$, 
where $\hat{\bm{n}}\times \equiv \epsilon_{ikj}\hat{n}_k \hat{\bm{e}}_i\otimes\hat{\bm{e}}_j$ and the interface plane is $\bm{x}\cdot\hat{\bm{n}} = c$ ($c$ is a constant).  
The total Burgers vector of all disconnections intersecting an arbitrary (probe) vector in the interface $\bm{p}$ is 
\begin{equation}\label{oFBE}
\bm{b}(\bm{p})
= 
\int_{\mathcal{S}} \bm{\upalpha}\hat{\bm{e}} \rmd a
= 
\llbracket \mathbf{F}^{-1} \rrbracket \bm{p},  
\end{equation}
where $\hat{\bm{e}} \equiv \hat{\bm{p}}\times \hat{\bm{n}}$, and $\mathcal{S}$ is an area normal to $\hat{\bm{e}}$ and its boundary spanning $\bm{p}$ (see Supplemental Material, SM~\cite{supp_mat}). 
Equation~\eqref{oFBE} is the classical FBE that describes interfaces containing only dislocations (without step character). 

Inclination of the interface plane in the natural state may be different from that in the reference state, associated with finite-height disconnection steps. 
Let $\hat{\bm{n}}_\txR$ be the normal vector of the reference interface. 
The step height density is
$\bm{\rho}
= (\hat{\bm{n}}_\txR \times \hat{\bm{n}}) 
\delta(\bm{x}\cdot\hat{\bm{n}} - c)$. 
The total step height of all disconnections intersecting $\bm{p}$ is 
\begin{equation}\label{ShpintSrhodot}
h(\bm{p}) 
= \int_{\mathcal{S}} 
\bm{\rho} \cdot \hat{\bm{e}} \rmd a
= \hat{\bm{n}}_\txR \cdot \bm{p}.
\end{equation}
A complete description of an interface with both mismatch in deformation/rotation and change in inclination is the combination of Eqs.~\eqref{oFBE} and \eqref{ShpintSrhodot}, i.e., GFBE: 
\begin{equation}\label{GFBE}
\underline{\bm{b}}(\bm{p})
= 
\underline{\mathbf{T}} \bm{p}
\quad
\forall \bm{p} \perp \hat{\bm{n}},
\end{equation}
where the extended Burgers-step vector and the incompatibility tensor are, respectively, defined as
\begin{equation}
\underline{\bm{b}}
\equiv
\left(\begin{array}{c}
\bm{b} \\ h
\end{array}\right)
\text{ and }
\underline{\mathbf{T}}
\equiv
\left(\begin{array}{c}
\llbracket \mathbf{F}^{-1} \rrbracket
\\
\hat{\bm{n}}_\txR^T
\end{array}\right). 
\end{equation}

We now interpret the total Burgers-step content in terms of discrete disconnection networks.
In the reference state, the coincidence-site lattice (CSL) and displacement-shift-complete (DSC) lattice~\cite{Sutton1983structure,Schwartz1985Atomic,Han2018Grain} determine the admissible disconnection modes: $\{\underline{\bm{b}}^{(m)} \equiv (\bm{b}^{(m)}, h^{(m)})\}$, with mode index $m$ ($= 1, \cdots, M$)~\cite{Admal2022interface}. 
For the $m^\text{th}$ set of parallel disconnections with line direction $\hat{\bm{\xi}}^{(m)}$ and spacing $d^{(m)}$, define the line density vector $\bm{\eta}^{(m)} \equiv \hat{\bm{n}} \times \hat{\bm{\xi}}^{(m)}/d^{(m)}$. 
The number of these disconnection lines intersecting $\bm{p}$ is $\bm{\eta}^{(m)} \cdot \bm{p}$.
Hence, the total Burgers-step content is 
\begin{equation}\label{Stbpsummgmdptbm}
\underline{\bm{b}}(\bm{p})
= 
\sum_{m=1}^M (\bm{\eta}^{(m)} \cdot \bm{p})
\underline{\bm{b}}^{(m)}
\quad
\forall \bm{p} \perp \hat{\bm{n}}. 
\end{equation}
Combining Eqs.~\eqref{GFBE} and \eqref{Stbpsummgmdptbm} yields the quantized GFBE (qGFBE) (see derivations in Sec.~I D 
of the SM~\cite{supp_mat}):  
\begin{equation}\label{qGFBE}
\underline{\mathbf{B}} \mathbf{N}^T
=
\underline{\mathbf{T}}^\parallel, 
\end{equation}
where
$\underline{\mathbf{B}}
\equiv
(\begin{array}{ccc}
\underline{\bm{b}}^{(1)} & \cdots & \underline{\bm{b}}^{(M)} 
\end{array})$,
$\mathbf{N}
\equiv
(\begin{array}{ccc}
\bm{\eta}^{(1)} & \cdots & \bm{\eta}^{(M)} 
\end{array})$,
and
$\underline{\mathbf{T}}^\parallel
\equiv
\underline{\mathbf{T}}(\mathbf{I} - \hat{\bm{n}} \otimes \hat{\bm{n}})$.
The qGFBE presents a system of linear equations for the unknown $\mathbf{N}$. 
A unique solution exists only when $\text{rank}([ \begin{array}{c|c} \underline{\mathbf{B}} & \underline{\mathbf{T}}^\parallel \end{array} ]) = \text{rank}(\underline{\mathbf{B}}) = M$; its closed form is $\mathbf{N} = (\underline{\mathbf{T}}^\parallel)^T (\underline{\mathbf{B}}^T)^*$, where ``$*$'' denotes the Moore-Penrose pseudoinverse. 

An interface structure is predicted as follows.
First, construct a reference state from the natural state by applying $(\mathbf{F}^{+/-})^{-1}$ on the adjoining crystals and specifying the reference interface normal $\hat{\bm{n}}_\txR$. 
This construction is not unique; a good choice is one for which (i) the reference plane is a close-packed plane in the CSL and (ii) both the norm of $\llbracket\mathbf{F}^{-1}\rrbracket$ and the angle $\arccos(\hat{\bm{n}} \cdot \hat{\bm{n}}_\text{R})$ are minimal. 
Second, enumerate admissible disconnection modes by the Smith normal crystallography method~\cite{Admal2022interface}. 
Third, starting from a disconnection mode with small $|\underline{\bm{b}}^{(m)}|$, solve the qGFBE. 
If the system is overdetermined, include additional disconnection modes or allow the inclination and/or misorientation to vary. 
Below, we illustrate applications of the qGFBE. 

\begin{table}[tb]
\caption{\label{tab:modes}
Summary of disconnection modes $\{\underline{\bm{b}}^{(m)}\}$ defined in the reference state. 
$a$ is the equilibrium lattice parameter.
$\{\hat{\bm{e}}_i\}$ are  coordinate axes attached to the reference interface plane; $\hat{\bm{e}}_3 \parallel \hat{\bm{n}}_\txR$.
}
\begin{tabular*}{\columnwidth}{@{\extracolsep{\fill}} c c c c @{}}
\toprule
Case &
$\hat{\bm{e}}_1 \times \hat{\bm{e}}_2 \times \hat{\bm{e}}_3$ &
$m$ &
$(\underline{\bm{b}}^{(m)}/a)^T$
\\ \midrule
\multirow{2}{*}{(i)}  & 
\multirow{2}{*}{$[10\bar{1}] \times [\bar{1}2\bar{1}] \times [111]$} & 
1 & 
$(\begin{array}{cccc} 0, & 0,  & \sqrt{3}/3, & \sqrt{3}/2\end{array})$
\\
&
&
2 &
$(\begin{array}{cccc} 0, & 0,  & \sqrt{3}/3, & -\sqrt{3}/2\end{array})$
\\ \midrule
\multirow{4}{*}{(ii)}  & 
\multirow{4}{*}{\shortstack[c]{$[1\bar{3}0] \times [310] \times [001]$ \\ for $\Sigma_\text{Au}:\Sigma_\text{Pd}=9:10$}} & 
1 & 
$(\begin{array}{cccc} \sqrt{10}/20, & 0,  & 0, & 1/2\end{array})$
\\
&
&
2 &
$(\begin{array}{cccc} \sqrt{10}/20, & 0,  & 0, & -1/2\end{array})$
\\
&
&
3 &
$(\begin{array}{cccc} 0, & \sqrt{10}/20,  & 0, & 1/2\end{array})$
\\
&
&
4 &
$(\begin{array}{cccc} 0, & \sqrt{10}/20,  & 0, & -1/2\end{array})$
\\ \midrule
\multirow{2}{*}{(iii)}  & 
\multirow{2}{*}{\shortstack[c]{$[110] \times [\bar{1}11] \times [1\bar{1}2]$ \\ referenced to $\upbeta$ phase}} & 
1 & 
$(\begin{array}{cccc} 0, & \sqrt{3}/12, & 0, & \sqrt{6}/3\end{array})$
\\
&
&
2 &
$(\begin{array}{cccc} \sqrt{2}/2, & \sqrt{3}/6, & \sqrt{6}/6, & 0\end{array})$
\\ \midrule
\multirow{3}{*}{(iv)}  & 
\multirow{3}{*}{\shortstack[c]{$[00\bar{1}] \times [110] \times [1\bar{1}0]$ \\ referenced to $\upbeta$ phase}} & 
1 & 
$(\begin{array}{cccc} 1/2, & -\sqrt{2}/2, & 0, & 0\end{array})$
\\
&
&
2 &
$(\begin{array}{cccc} 1, & 0, & 0, & 0\end{array})$
\\
&
&
3 &
$(\begin{array}{cccc} 0, & 0, & 0, & \sqrt{2}\end{array})$
\\ \bottomrule
\end{tabular*}
\end{table}

\begin{figure}[tb]
\includegraphics[width=0.98\linewidth]{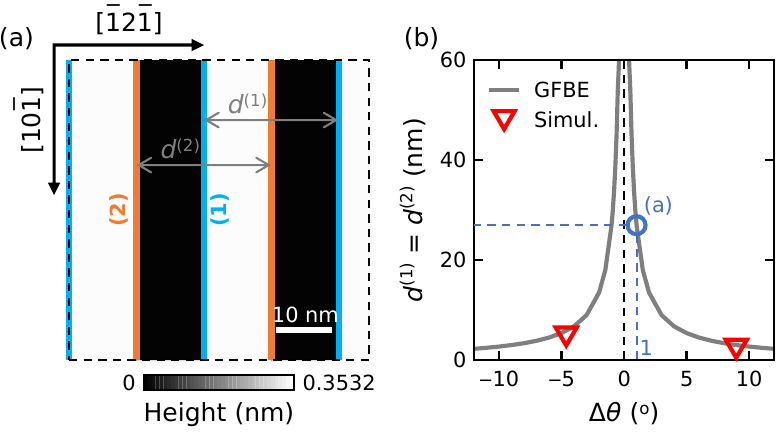}
\caption{\label{twin} 
Structure of a misoriented coherent twin boundary. 
(a) Plan view of interface structure for $\Delta\theta = 1^\circ$. 
blue and orange lines denote mode-(1) and (2) disconnections. 
(b) The spacing between disconnections of the same mode ($d^{(1)} = d^{(2)}$) vs. the deviation from the $\Sigma 3$ misorientation ($\Delta\theta$). 
The solid curve is the qGFBE prediction. 
Red triangles and blue circle denote, respectively, atomistic simulation results~\cite{OBrien2016Misoriented} and the case shown in (a). 
} 
\end{figure}

{\it Case (i): Misoriented coherent twin boundaries}---The $\Sigma 3$ $(111)$ $\langle 1\bar{1}0 \rangle$ coherent twin boundary (CTB) is one of the most important (and widely studied) interfaces in face-centered cubic (FCC) metals. 
A small deviation from an ideal $\Sigma 3$ misorientation produces a misoriented CTB~\cite{marquis2005structural,OBrien2016Misoriented}.
We take the ideal CTB as the reference state and denote the additional misorientation by $\Delta \theta$. 
The interface inclination remains unchanged ($\hat{\bm{n}} = \hat{\bm{n}}_\txR$) while $\mathbf{F}^{+/-}$ corresponds to rotations of the two grains by $\pm \Delta\theta/2$ about the tilt axis $[10\bar{1}]$. 
The admissible disconnection modes with short Burgers-step vectors are listed in Table~\ref{tab:modes}. 
Solving the qGFBE yields two sets of disconnections parallel to the tilt axis with equal spacings $d^{(1)} = d^{(2)}$ (Fig.~\ref{twin}a); the relative positions of the two sets are undetermined by the qGFBE. 
(Minimizing the elastic energy suggests the relative spacing of $d^{(1)}/2$.) 
Figure~\ref{twin}b shows the predicted spacing between disconnections of the same mode vs. $\Delta\theta$. 
The result reduces to Frank's formula~\cite{frank1950report} and agrees with the atomistic simulation results for Ni~\cite{OBrien2016Misoriented} (red triangles). 
Detailed analysis is given in Sec.~II 
of the SM~\cite{supp_mat}.

\begin{figure}[tb]
\includegraphics[width=0.98\linewidth]{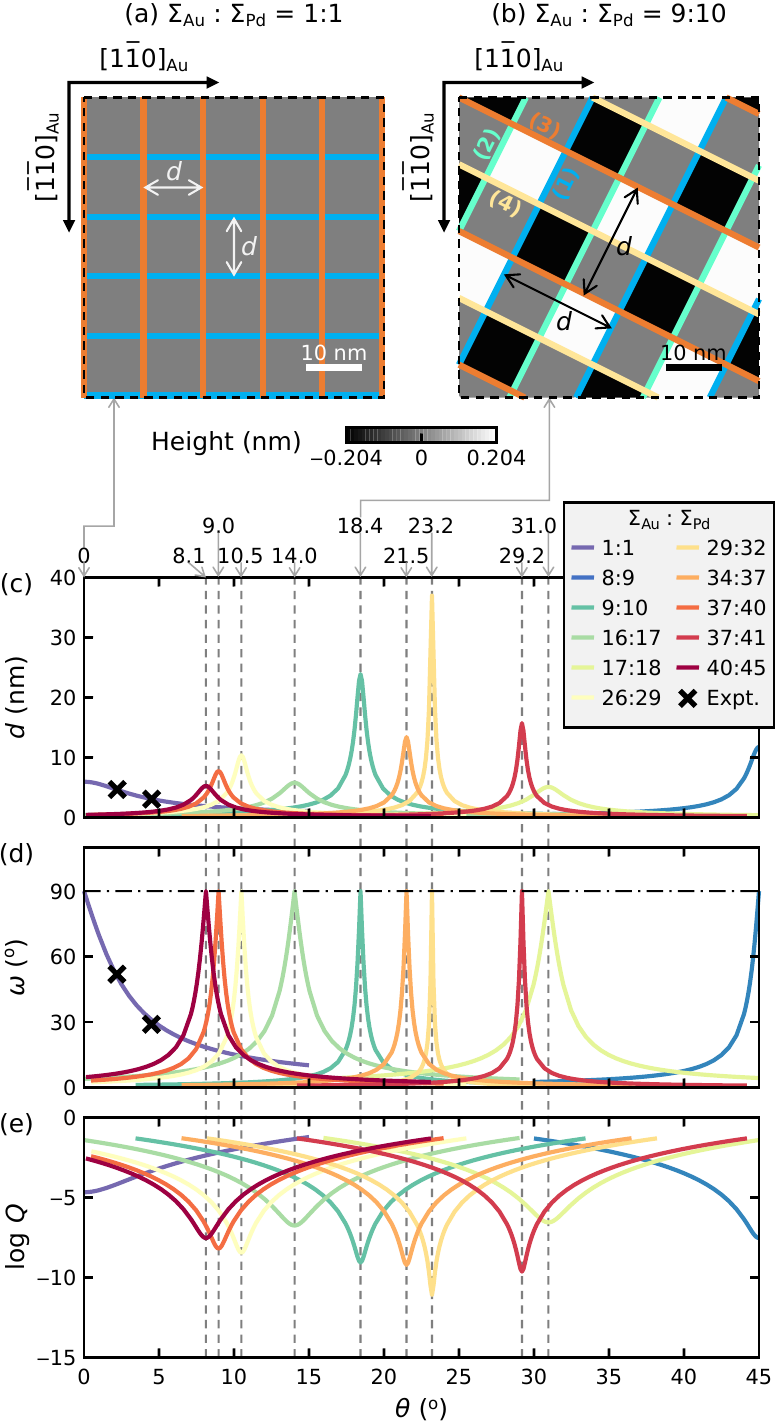}
\caption{\label{twisthetero} 
Structure of twist Au-Pd heterophase interfaces. 
(a, b) Interface structures for $\Sigma_\text{Au}:\Sigma_\text{Pd}=1:1$ ($\theta=0$) and $9:10$ ($\theta\approx 18.4^\circ$).
In (a), blue and orange lines denote mode-(1) and (2) disconnections;
In (b), blue, green, orange, and yellow lines denote mode-(1) to (4) disconnections (Table~\ref{tab:modes}). 
The background color indicates the interface height. 
(c-e) Disconnection line spacing ($d$), the angle between Burgers vector and line direction ($\omega$), and $\log Q$ (defined in the text) vs. misorientation ($\theta$). 
Colors indicate different reference states. 
The experimental data~\cite{Hwang1980Interphase} represented by the black crosses are consistent with the $1:1$ reference state prediction.
} 
\end{figure}

{\it Case (ii): Twist heterophase interfaces}---Consider an interface formed by joining two FCC crystals along their $(001)$ planes with the $[100]$ and $[010]$ directions initially aligned, followed by a relative rotation about $[001]$ by an angle $\theta$. 
Such twist heterophase interfaces have been experimentally studied in Au-Pd~\cite{Hwang1980Interphase}. 
At certain twist angles, coherent interfaces can be obtained by applying small in-plane strains~\cite{Hwang1980Interphase, Balluffi1982CSLDSC}; the cases considered here require strains $<0.02$, indicated by the vertical dashed lines in Figs.~\ref{twisthetero}c-e.
For each coherent interface, $\Sigma_\text{Au/Pd}$ denotes the reciprocal density of coincidence sites referenced to the Au/Pd lattice; the ratio $\Sigma_\text{Au}:\Sigma_\text{Pd}$ specifies the matching relation between the two $(001)$ planes. 
For example, the coherent interface at $\theta = 0$ corresponds to $\Sigma_\text{Au}:\Sigma_\text{Pd}=1:1$ (Fig.~\ref{twisthetero}a); using it as the reference state, the qGFBE predicts two sets of edge disconnections with zero step height. 
In contrast, other coherent interfaces, e.g., $\Sigma_\text{Au}:\Sigma_\text{Pd}=9:10$ (Fig.~\ref{twisthetero}b), require four sets of edge disconnections with finite step heights (Table~\ref{tab:modes}), a feature not captured within the classical FBE.

A deviation from any coherent misorientation introduces screw character into the disconnections.
Figures~\ref{twisthetero}c, d show the disconnection line spacing ($d$) and the angle between the Burgers vector and line direction ($\omega$) vs. $\theta$; colors indicate the predictions using different reference states.
Each coherent misorientation corresponds to a local maximum in $d$ and $\omega = 90^\circ$ (i.e., pure edge disconnections). 
Detailed analysis is given in Sec.~III 
of the SM~\cite{supp_mat}.
As a crude measure of disconnection network energy, we define $Q \equiv \sum_{m,n} (b^{(m)}b^{(n)})/(d^{(m)}d^{(n)})$~\cite{Ecob1980Geometrical, Vattre2014Computational}; Fig.~\ref{twisthetero}e shows $\log Q$ vs. $\theta$. 
Unlike pure twist grain boundaries, the qGFBE predicts no cusp at coherent misorientations, including $\theta=0$ (i.e., $\log Q(\theta)$ is differentiable everywhere). 
Moreover, $\theta=0$ is not the lowest-energy interface. 
Some finite-angle coherent interfaces, such as $\Sigma_\text{Au}:\Sigma_\text{Pd}=29:32$, exhibit lower $Q$ because they require less coherency strains and hence larger disconnection spacings. 
(Note: a complete interface energy estimate should also include the energies of the coherent reference state and the disconnection steps.)

\begin{figure*}[tb]
\includegraphics[width=0.98\textwidth]{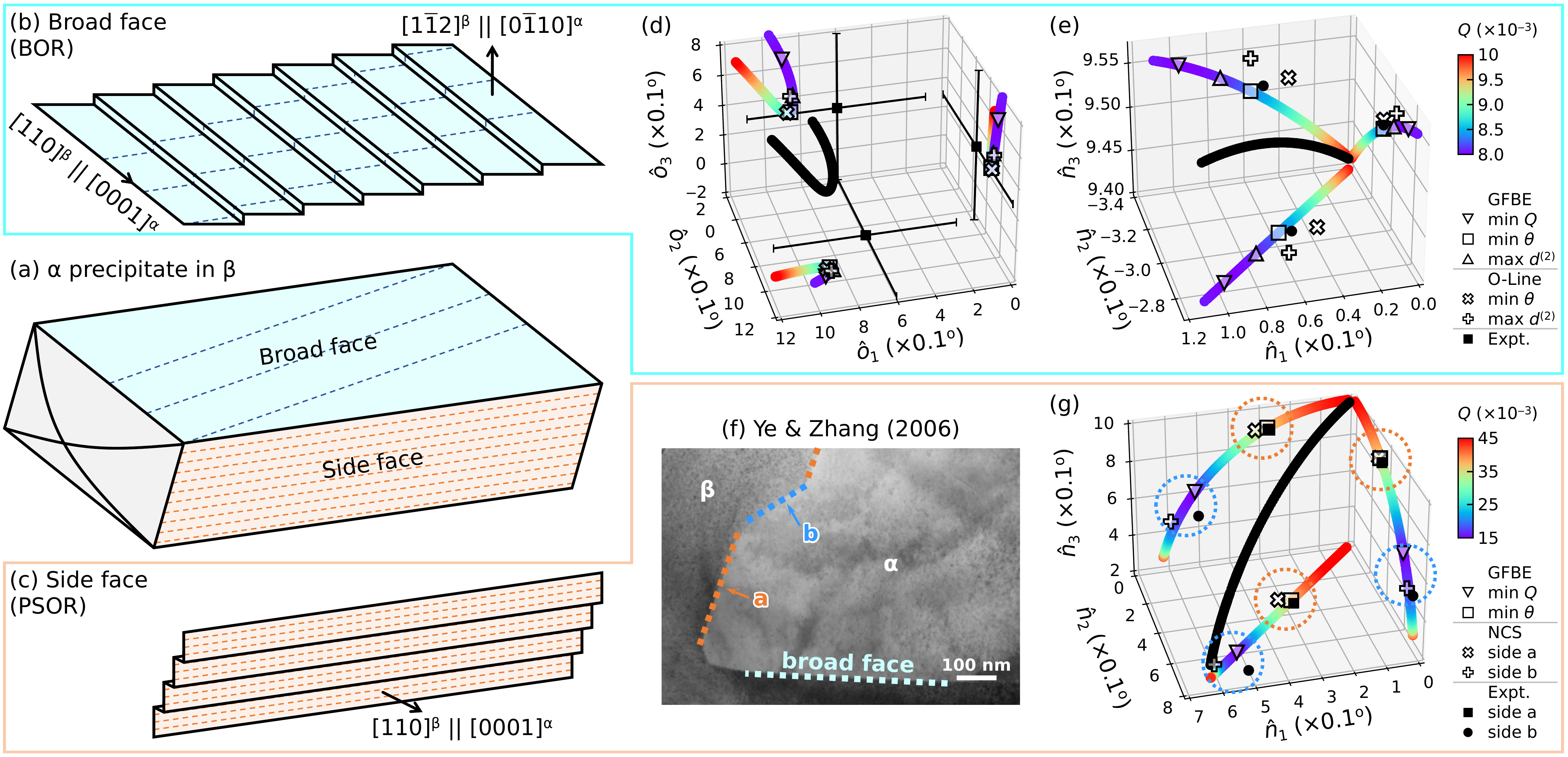}
\caption{\label{precipitate} 
(a) Schematic of an $\upalpha$ nucleus in $\upbeta$ phase. 
(b) Broad face; steps and dashed lines denote mode-(1) and (2) disconnections. 
(c) Side face; dashed lines denote mode-(1) and (2) disconnections, and steps denote mode-(3).  
(d, e) qGFBE solution (black curves) projected onto the misorientation and inclination spaces, respectively. 
$\hat{o}_1$, $\hat{o}_2$, and $\hat{o}_3$ are the angles between $[0001]^\upalpha$ and $[110]^\upbeta$, $[0\bar{1}10]^\upalpha$ and $[1\bar{1}2]^\upbeta$, and $[\bar{2}110]^\upalpha$ and $[\bar{1}11]^\upbeta$, respectively. 
$\{\hat{n}_i\}$ are components of the interface normal $\hat{\bm{n}}$. 
The projections of the black curves onto the coordinate planes are shown in color, colored by $Q$.
Symbols (see legend) denote qGFBE predictions, O-line predictions~\cite{Qiu2013Crystallography}, and experiments~\cite{Qiu2013Crystallography}.  
(f) Two side-face facets, ``a'' and ``b'', observed experimentally~\cite{Ye2006Dislocation}. 
(g) qGFBE solution projected onto the inclination space (black curve) and its projection onto the coordinate planes colored by $Q$. 
Symbols (see legend) denote qGFBE predictions, near-coincidence-site (NCS) predictions~\cite{Ye2006Near}, and experiments~\cite{Ye2006Dislocation}. 
Orange and blue dotted circles mark facets ``a'' and ``b'' in (f), respectively.
} 
\end{figure*}

{\it Case (iii): Broad face of a martensite nucleus}---The high-temperature phase of Ti is body-centered cubic (BCC) $\upbeta$. Upon cooling, hexagonal close-packed (HCP) $\upalpha$ nucleates within $\upbeta$. 
The $\alpha$ nucleus is lath-shaped; its broad face (Figs.~\ref{precipitate}a, b) lies close to the coherent $(1\bar{1}2)^\upbeta \parallel (0\bar{1}10)^\upalpha$ interface satisfying the Burgers orientation relationship (BOR)~\cite{Furuhara1991interface, Miyano2002HRTEM, Zhang2021Structures}. 
We treat the coherent interface, formed by straining the $\upalpha$ lattice to match $\upbeta$, as the reference state.

We first neglect the mismatch along $[110]^\upbeta \parallel [0001]^\upalpha$ so that an invariant plane exists\cite{Pond2003comparison}, and assume a single disconnection mode. 
The experimentally observed mode is mode-(1) in Table~\ref{tab:modes}~\cite{Ackerman2020PRM}.
In this case, the qGFBE (Eq.~\eqref{qGFBE}) is overdetermined because $\text{rank}([\begin{array}{c|c} \underline{\mathbf{B}} & \underline{\mathbf{T}}^\parallel \end{array}]) > \text{rank}(\underline{\mathbf{B}})$. 
We relax the interface inclination ($\in \mathrm{S}^2$) and misorientation ($\in \mathrm{SO}(3)$) away from BOR. 
The resulting solution suggests that the disconnection line, inclination axis, and misorientation axis are all parallel to $[110]^\upbeta$ (see Sec.~IV A 
of the SM~\cite{supp_mat}). 
The predicted line spacing $d$, inclination angle $\phi$, and misorientation angle $\theta$ for Ti-6246 and Mg-Li alloys are listed in Table~\ref{tablePTMCTM}. 
The qGFBE predictions agree well with the phenomenological theory of martensite crystallography (PTMC)~\cite{Wechsler1953PTMC, BOWLES1954PTMC}, the topological model (TM)~\cite{POND1994Defects, Pond2003comparison}, and experiments~\cite{Ackerman2020PRM, Kral2007Crystallography}. 
The qGFBE predictions coincide with those from the PTMC. 
But, unlike PTMC, the qGFBE also predicts the disconnection network, and remains applicable when no invariant plane exists (as is generally the case). 
Approximations made in TM predictions lead to slight deviations from the qGFBE and PTMC results. 
In addition, the qGBFE is easier to automate than TM (which requires skillful geometric analysis). 

\begin{table}[tb]
\caption{\label{tablePTMCTM}
The misorientation angle ($\theta$), inclination angle ($\phi$), and line spacing of mode-(1) disconnections ($d^{(1)}$) of the $\upalpha$/$\upbeta$ interface obtained by different methods.
} 
\begin{tabular*}{\columnwidth}{@{\extracolsep{\fill}} c c c c c @{}}
\toprule
~ & qGFBE & PTMC & TM & Experiment \\
\midrule
Ti-6246 \\ [5pt]
$\theta$ ($^{\circ}$) & $0.5331$ & $0.5331$ & $0.5336$ & -- \\
$\phi$ ($^{\circ}$)   & $12.267$ & $12.267$ & $12.290$ & $12.6\pm 1.0$~\cite{Ackerman2020PRM} \\
$d^{(1)}$ (nm)              & $1.250$  & --       & $1.247$  & $1.46\pm 0.37$~\cite{Ackerman2020PRM} \\
\midrule
Mg-Li \\ [5pt]
$\theta$ ($^{\circ}$) & $0.5238$ & $0.5238$ & $0.5242$ & $0.6\pm 0.1$~\cite{Kral2007Crystallography} \\
$\phi$ ($^{\circ}$)   & $14.291$ & $14.291$ & $14.318$ & $13.7\pm 0.5$~\cite{Kral2007Crystallography} \\
$d^{(1)}$ (nm)              & $1.162$  & --       & $1.160$  & -- \\
\bottomrule
\end{tabular*}
\end{table}

To include the actual lattice mismatch along $[110]^\upbeta \parallel [0001]^\upalpha$, we next consider both mode-(1) and (2) disconnections (Table~\ref{tab:modes}), consistent with experimental observations~\cite{zheng2018determination}. 
Again, the interface inclination and misorientation are allowed to vary. 
The compatibility condition $\text{rank}([\begin{array}{c|c} \underline{\mathbf{B}} & \underline{\mathbf{T}}^\parallel \end{array}]) = \text{rank}(\underline{\mathbf{B}}) = 2$ defines a one-dimensional solution manifold in $\mathrm{S}^2 \times \mathrm{SO}(3)$, i.e., a curve in the five-dimensional interface normal-misorientation space (see Sec.~IV B 
of the SM~\cite{supp_mat}).
Figures~\ref{precipitate}d and e show projections of this curve onto the misorientation and inclination spaces. 
An additional criterion is needed to select a point on this manifold. 
Without addressing the merits of different criteria, we consider: minimizing the disconnection energy measure $Q$ ($\min Q$), minimizing the deviation from the BOR ($\min\theta$), and maximizing the spacing of mode-(2) disconnections ($\max d^{(2)}$)~\footnote{$Q$ provides a crude measure of the disconnection-network energy. $\min\theta$ maximizes compatibility with the misorientation on other facets of the $\alpha$ nucleus. $\max d^{(2)}$ applies when mode-(2) disconnections dominate the energy.}. 
The corresponding solutions are shown in Figs.~\ref{precipitate}d, e. 
Among them, the $\min\theta$ solution agrees best with experiment (open squares vs. solid circles in Fig.~\ref{precipitate}e).

{\it Case (iv): Side face of a martensite nucleus}---The side face of the $\upalpha$ nucleus in a $\upbeta$ Ti alloy is illustrated in Figs.~\ref{precipitate}a, c. 
The reference state is constructed by straining the $\upalpha$ phase to match $\upbeta$ coherently along the $(1\bar{1}0)^\upbeta \parallel (1\bar{1}00)^\upalpha$ plane with the Pitsch-Schrader orientation relationship (PSOR)~\cite{Pitsch1958Ausscheidungsform, Dahmen1982Orientation}. 
The admissible disconnection modes are listed in Table~\ref{tab:modes}. 
Mode-(1) and (2) disconnections are pure dislocations while mode-(3) is a pure step. 
In this case, the qGFBE is underdetermined.
We assume that the three sets of disconnections are parallel, as suggested by the experimental observations and atomistic simulation results~\cite{Furuhara1991interface, Miyano2002HRTEM, Ye2006Dislocation, Ye2006Near, Zhang2021Structures}. 
With this assumption, the solution is a curve in the inclination-misorientation space (see Sec.~IV C 
of the SM~\cite{supp_mat}), shown as its projection onto the inclination space Fig.~\ref{precipitate}g.
We also show the points corresponding to $\min Q$ and $\min \theta$. 
In the experiments of Ye and Zhang~\cite{Ye2006Dislocation}, two types of side facets are observed (denoted  ``a'' and ``b'' in Fig.~\ref{precipitate}f). 
From Fig.~\ref{precipitate}g, we see that the solution corresponding to $\min\theta$ predicts that facet ``a'' is favorable while the solution to $\min Q$ predicts facet ``b''.

The current analytical GFBE-based method cannot accurately describe detailed interface structure when disconnection reactions occur (e.g., the formation of the honeycomb dislocation network observed in twist grain boundaries). 
However, the GFBE can be used to generate an initial prediction of the disconnection network that can be further relaxed through either atomistic simulations, microscopic phase field models~\cite{shen2014predicting}, or the generalized Peierls-Nabarro model~\cite{Xiang2008generalizedPN} approaches.

\begin{acknowledgments}
JH acknowledges support from the General Research Fund of the Research Grants Council of Hong Kong (11213425); 
JH, DSJ, and YX acknowledge support from the Collaborative Research Fund of the Research Grants Council of Hong Kong (C7140-25GF); 
DSJ acknowledges support from the General Research Fund of the Research Grants Council of Hong Kong (17210723);
LZ acknowledges support from the National Natural Science Foundation of China (12571464, 12201423, and 12426311), the Guangdong Key Laboratory of Applied Mathematics and Artificial Intelligence, Shenzhen Science and Technology Innovation Program (RCYX20231211090222026, JCYJ20241202124209011), and the Research Team Cultivation Program of Shenzhen University (2023QNT011).
\end{acknowledgments}

\bibliography{GFBE}

\end{document}


\title{{\sc supplemental material}\\
A Generalized Frank-Bilby Equation for Interfaces in Crystalline Materials}

\author{Dongsong Tao}
\affiliation{Department of Materials Science and Engineering, City University of Hong Kong, Hong Kong SAR, China}
\author{Luchan Zhang}
\affiliation{College of Mathematics and Statistics, Shenzhen University, Shenzhen, China}
\author{David J. Srolovitz}
\affiliation{Department of Mechanical Engineering, The University of Hong Kong, Hong Kong SAR, China}
\author{Yang Xiang}
\affiliation{Department of Mathematics, The Hong Kong University of Science and Technology, Hong Kong SAR, China}
\author{Jian Han}
\email{jianhan@cityu.edu.hk}
\affiliation{Department of Materials Science and Engineering, City University of Hong Kong, Hong Kong SAR, China}


\maketitle

\tableofcontents

\section{Derivation of the generalized Frank-Bilby equation}

\subsection{Closure failure of Burgers circuit across an interface}

\begin{figure}[tb]
\begin{center}
\includegraphics[width=0.85\columnwidth]{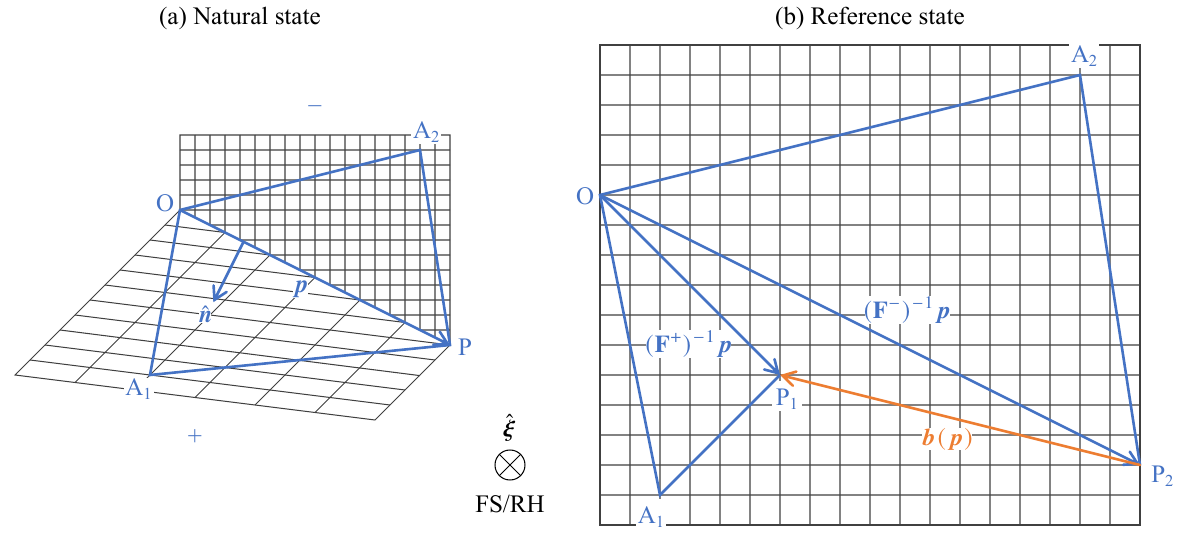}
\caption{\label{FB}
Determination of the total Burgers vector in an interface by Burgers circuits. 
A closed, right-handed circuit PA$_1$OA$_2$P enclosing the probe vector $\bm{p}$ in the natural state (a) is mapped onto the path P$_1$A$_1$OA$_2$P$_2$ in the reference state (b) with closure failure P$_2$P$_1$.
FS/RH stands for finish-to-start right-handed convention.  
$\hat{\bm{\xi}} = \hat{\bm{p}} \times \hat{\bm{n}}$ is the direction for drawing the right-handed circuits.
} 
\end{center}
\end{figure}

Consider two lattices that meet at a flat interface; such configuration is referred to as the natural state (NS). 
In general, the lattices may differ in Bravais lattice type and/or orientation. 
As illustrated in Fig.~\ref{FB}, the two lattices in NS can be viewed as arising from a reference state (RS) via affine transformations $\mathbf{F}^+$ and $\mathbf{F}^-$, where ``$+$''/``$-$'' denotes the lattice in the half space pointed by the interface normal vector $+\hat{\bm{n}}$/$-\hat{\bm{n}}$. 
NS and RS may each represent a single-crystal lattice or a displacement-shift-complete (DSC) lattice. 
Since only their translational symmetries are relevant, any rigid translation or the actual atomic arrangement in NS can be disregarded. 
Without loss of generality, we shift the two lattices in NS to share a common origin O; see Fig.~\ref{FB}a. 

In NS, let $\text{OP} = \bm{p}$ be a probe vector lying in the interface. 
The total Burgers vector intersecting $\bm{p}$ can be obtained from the closure failure of a Burgers circuit enclosing $\bm{p}$. 
Following the finish-to-start/right-handed (FS/RH) convention, 
\begin{itemize}
\item[(i)]
Define $\hat{\bm{\xi}} = \hat{\bm{p}} \times \hat{\bm{n}}$. 
\item[(ii)]
Construct a closed, right-handed circuit PA$_1$OA$_2$P about $\hat{\bm{\xi}}$ in NS.  
\item[(iii)]
Map the path into RS using $(\mathbf{F}^+)^{-1}$ and $(\mathbf{F}^-)^{-1}$, and measure the closure failure as the vector from the finish to the start point. 
\end{itemize}
The segment PA$_1$O in the ``$+$'' lattice of NS is deformed to P$_1$A$_1$O in RS, while the segment OA$_2$P in the ``$-$'' lattice of NS is deformed to OA$_2$P$_2$ in RS. 
The closure failure, defined as a vector from P$_2$ to P$_1$, gives the total Burgers vector crossing $\bm{p}$: 
\begin{equation}\label{FBeq}
\bm{b}(\bm{p}) = \text{P}_2\text{P}_1 = \text{O}\text{P}_1 - \text{O}\text{P}_2
= 
\left[(\mathbf{F}^+)^{-1} - (\mathbf{F}^-)^{-1}\right] \bm{p}.
\end{equation}
This equation, first derived by Frank~\cite{frank1950report} for grain boundaries (GBs) and later generalized to heterophase interfaces by Bilby~\cite{bilby1955report}, is known as the classical \emph{Frank-Bilby equation (FBE)}. 

\subsection{The classical Frank-Bilby equation}

Before considering interfaces, we begin by recalling the essential continuum kinematics of single crystals. 
The displacement gradient $\mathbf{H} = \uGrad \mathbf{u}$ is additively decomposed into elastic and plastic distortions as
\begin{equation}\label{Hdecompose}
\mathbf{H} 
= \uGrad\bm{u}
= \mathbf{H}^\txe + \mathbf{H}^\txp. 
\end{equation}
For any smooth vector field, the curl of its gradient vanishes; thus, for
a closed loop $\rmC$ that bounds a smooth oriented area $\rmS$ in RS, we have
\begin{equation}\label{ointCHdXointCGradudXintSCurlGraduTedA0}
\oint_\rmC \mathbf{H} \rmd\bm{X}
= \oint_\rmC \uGrad\bm{u} \rmd\bm{X}
\xrightarrow{\text{Stokes' theorem}}
\int_\rmS \left(\uCurl\uGrad\bm{u}\right)^T \hat{\bm{e}} \rmd A
= \mathbf{0}, 
\end{equation}
where $\hat{\bm{e}}$ is the unit normal vector to the oriented area $\rmS$. 
In the presence of dislocations threading the area $\rmS$, the elastic part remains curl‑free, but the plastic part $\mathbf{H}^\txp$ gives a nonzero line integral. 
The total Burgers vector of all dislocation lines piercing $\rmS$ is then
\begin{equation}\label{bHp}
\bm{b}(\rmC)
= \oint_\rmC \mathbf{H}^\txp \rmd\bm{X}
\xrightarrow{\text{Stokes' theorem}}
\int_\rmS \left(\uCurl\mathbf{H}^\txp\right)^T \hat{\bm{e}} \rmd A
= \int_\rmS \mathbf{G}^T \hat{\bm{e}} \rmd A, 
\end{equation}
where we have introduced the Burgers tensor:
\begin{equation}\label{defBurgerstensor}
\mathbf{G}
\equiv \uCurl\mathbf{H}^\txp. 
\end{equation}
The quantity $\mathbf{G}^T \hat{\bm{e}}$ represents the Burgers vector per unit area of dislocation lines piercing a plane with normal $\hat{\bm{e}}$. 
Because $\mathbf{H}^\txp$, in general, cannot be expressed as the gradient of a vector field, the line integral Eq.~\eqref{bHp} is nonzero. 
From Eq.~\eqref{Hdecompose} and Eq.~\eqref{ointCHdXointCGradudXintSCurlGraduTedA0}, we also have
\begin{equation}
\mathbf{G} = - \uCurl\mathbf{H}^\txe, 
\end{equation}
which emphasizes that elastic and plastic incompatibilities are not independent; their curls must cancel to give a compatible total displacement gradient.

The additive decomposition Eq.~\eqref{Hdecompose} only works for small deformations. 
For finite (large) deformations, we should work with the deformation gradient $\mathbf{F}$ and use the multiplicative decomposition: 
\begin{equation}
\mathbf{F} = \mathbf{F}^\txe \mathbf{F}^\txp
\end{equation}
instead of decomposing $\mathbf{H}$. 
If we define elastic and plastic distortions as
\begin{equation}
\mathbf{H}^\txe = \mathbf{F}^\txe - \mathbf{I}
\quad\text{and}\quad 
\mathbf{H}^\txp = \mathbf{F}^\txp - \mathbf{I},
\end{equation}
then
\begin{equation}
\mathbf{F} 
= (\mathbf{H}^\txe + \mathbf{I})(\mathbf{H}^\txp + \mathbf{I})
= \mathbf{H}^\txe + \mathbf{H}^\txp + \mathbf{I} + \mathbf{H}^\txe\mathbf{H}^\txp
\xrightarrow{\text{keep linear terms}}
\approx \mathbf{H}^\txe + \mathbf{H}^\txp + \mathbf{I}
~\Rightarrow~
\mathbf{H}
\approx  
\mathbf{H}^\txe + \mathbf{H}^\txp.
\end{equation}
Thus, the additive decomposition Eq.~\eqref{Hdecompose} is recovered as the linearized form of the multiplicative decomposition, and is valid only when $\Vert\mathbf{H}^\txe\mathbf{H}^\txp\Vert$ is negligible.  
In the following, we extend these concepts to the case of an interface. 

\begin{figure}[tb]
\includegraphics[width = 0.65\linewidth]{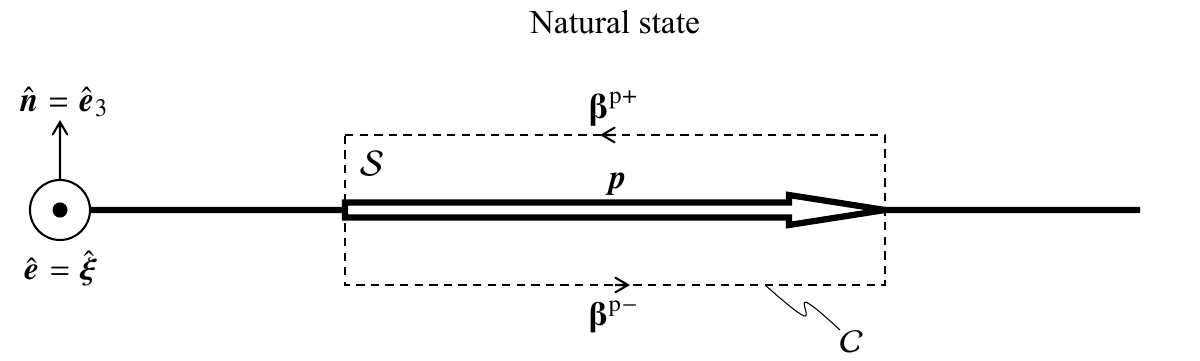}
\caption{\label{FBdensity}
Rectangular circuit $\mathcal{C}$ enclosing a part of interface in NS. 
$\hat{\bm{n}}$ is the interface normal. 
$\hat{\bm{e}}$ is the normal of the rectangular area and it is perpendicular to $\hat{\bm{n}}$. 
Across the interface the distortion tensor changes from $\bm{\upbeta}^{\txp -}$ below to $\bm{\upbeta}^{\txp +}$ above. 
} 
\end{figure}

Consider now the interface configuration shown schematically in Fig.~\ref{FB}. 
Let RS serve as the reference configuration, with coordinates denoted by $\bm{X}$, and NS as the deformed configuration, with coordinates denoted by $\bm{x}$. 
It is more convenient to work in NS, because the interface is explicitly present there. 
We choose a thin rectangular shell parallel to the interface as the Burgers circuit; see Fig.~\ref{FBdensity}. 
We cannot directly apply the integral Eq.~\eqref{bHp} since it is performed along a circuit in RS. 
We need to transform the integral in RS to an integral in NS, i.e., change the integral variable from $\bm{X}$ to $\bm{x}$. 
By chain rule, 
\begin{equation}
\mathbf{H}\label{Hbeta}
= \uGrad\bm{u}
= (\ugrad\bm{u})\mathbf{F}
= \bm{\upbeta}^T \mathbf{F}, 
\end{equation}
where we have introduced the distortion tensor $\bm{\upbeta} \equiv (\ugrad\bm{u})^T$. 
We assume that this distortion tensor can be additively decomposed into elastic and plastic parts:
\begin{equation}
\bm{\upbeta} 
= 
\bm{\upbeta}^\txe + \bm{\upbeta}^\txp. 
\end{equation}
Then, from Eqs.~\eqref{Hdecompose} and \eqref{Hbeta}, 
\begin{equation}\label{HpbetapF}
\mathbf{H}^\txe + \mathbf{H}^\txp
= \left(
{\bm{\upbeta}^\txe}^T 
+ 
{\bm{\upbeta}^\txp}^T
\right) \mathbf{F}
\quad\Rightarrow\quad
\mathbf{H}^\txe = {\bm{\upbeta}^\txe}^T \mathbf{F}
\text{ and }
\mathbf{H}^\txp = {\bm{\upbeta}^\txp}^T \mathbf{F}. 
\end{equation}
Now consider the total Burgers vector in Eq.~\eqref{bHp}. 
Using the relation $\rmd\bm{x} = \mathbf{F}\,\rmd\bm{X}$ along the Burgers circuit, we transform the integral from RS to NS:
\begin{equation}\label{bintCHpdXintCbpTFFinvdx}
\bm{b}
= \oint_\rmC \mathbf{H}^\txp \rmd \bm{X}
= \oint_{\mathcal{C}} \left({\bm{\upbeta}^\txp}^T\mathbf{F}\right) 
\left(\mathbf{F}^{-1}\rmd \bm{x}\right)
= \oint_{\mathcal{C}} {\bm{\upbeta}^\txp}^T \rmd \bm{x}
\xrightarrow{\text{Stokes' theorem}}
\int_{\mathcal{S}} \left(\ucurl{\bm{\upbeta}^\txp}^T\right)^T \hat{\bm{e}} \rmd a
\equiv
\int_{\mathcal{S}} \bm{\upalpha} 
\hat{\bm{e}} \rmd a, 
\end{equation}
where 
\begin{equation}\label{defNyetensor}
\bm{\upalpha}
\equiv 
\left(\ucurl{\bm{\upbeta}^\txp}^T\right)^T
\end{equation}
is the Nye tensor. 
Comparing Eq.~\eqref{defNyetensor} with Eq.~\eqref{defBurgerstensor} shows that the Nye tensor and Burgers tensor are closely related but not identical: $\bm{\upalpha}$ is defined in NS, whereas $\mathbf{G}$ is defined in RS.
To proceed, we write Eq.~\eqref{defNyetensor} in components:
\begin{equation}\label{curlbetapTeipqbetapTjqpeiejeipq}
\ucurl{\bm{\upbeta}^\txp}^T
= e_{ipq} [{\bm{\upbeta}^\txp}^T]_{jq,p}
\hat{\bm{e}}_i \otimes \hat{\bm{e}}_j
= e_{ipq} \beta^\txp_{qj,p}
\hat{\bm{e}}_i \otimes \hat{\bm{e}}_j
\quad\Rightarrow\quad
\bm{\upalpha}
= 
e_{ipq} \beta^\txp_{qj,p}
\hat{\bm{e}}_j \otimes \hat{\bm{e}}_i. 
\end{equation}
Next, we specialize to the planar interface configuration shown in Fig. ~\ref{FBdensity} and choose the basis such that $\hat{\bm{e}}_3$ is the unit normal $\hat{\bm{n}}$ to the interface. 
The plastic distortion varies only in the direction normal to the interface, so that $\beta^\txp_{qj,p}\ne 0$ only when $p = 3$. 
Equation~\eqref{curlbetapTeipqbetapTjqpeiejeipq} then reduces to
\begin{equation}\label{alphaeitqbpqjeejoeiejom}
\bm{\upalpha}
= 
e_{i3q} \beta^\txp_{qj,3}
\hat{\bm{e}}_j \otimes \hat{\bm{e}}_i
= 
\hat{\bm{e}}_j \otimes 
\left(-\beta^\txp_{2j,3}\hat{\bm{e}}_1 
+
\beta^\txp_{1j,3}\hat{\bm{e}}_2\right)
= 
-\hat{\bm{e}}_j \otimes 
\left(\beta^\txp_{ij,3}\hat{\bm{e}}_i \times\hat{\bm{e}}_3\right)
= 
-\left(\hat{\bm{e}}_j \otimes \beta^\txp_{ij,3}\hat{\bm{e}}_i\right) 
\left(\hat{\bm{e}}_3\times\right). 
\end{equation}
The last step follows from the identity: 
$\bm{u}\otimes(\bm{v}\times\bm{w}) = (\bm{u}\otimes\bm{v})(\bm{w}\times)$
for any vectors $\bm{u}$, $\bm{v}$, and $\bm{w}$. 
The plastic distortion within each crystal should be piecewise constant, i.e., $\bm{\upbeta}^{\txp+}$ and $\bm{\upbeta}^{\txp-}$ are not functions of $\bm{x}$. 
The jump across the interface is denoted by $\llbracket\bm{\upbeta}^\txp\rrbracket \equiv \bm{\upbeta}^{\txp+} - \bm{\upbeta}^{\txp-}$. 
The plastic distortion field can then be written as
\begin{equation}\label{betapxjumpbetapTheta}
\bm{\upbeta}^\txp(\bm{x})
= \llbracket\bm{\upbeta}^\txp\rrbracket \Theta(x_3)
+ \bm{\upbeta}^{\txp-}, 
\end{equation}
where $\Theta(x)$ is the Heaviside step function. 
Differentiating with respect to $x_3$​ yields
\begin{equation}\label{betapqjxjumpbpqjdx}
\beta^\txp_{ij,3}(\bm{x})
= \llbracket \beta^\txp_{ij} \rrbracket \delta(x_3). 
\end{equation}
Substituting Eq.~\eqref{betapqjxjumpbpqjdx} into Eq.~\eqref{alphaeitqbpqjeejoeiejom}, we obtain
\begin{equation}\label{alphambpijejoeietcrossdxt}
\bm{\upalpha}
= 
-\llbracket\beta^\txp_{ij}\rrbracket 
(\hat{\bm{e}}_j \otimes \hat{\bm{e}}_i) 
(\hat{\bm{e}}_3\times) \delta(x_3)
=
-\llbracket{\bm{\upbeta}^\txp}^T\rrbracket 
(\hat{\bm{n}}\times) \delta(x_3). 
\end{equation}
Next, we relate the plastic distortion $\bm{\upbeta}^\txp$ to the deformation gradient $\mathbf{F}$. 
From Eq.~\eqref{HpbetapF}, 
\begin{equation}
{\bm{\upbeta}^\txp}^T
=
\mathbf{H}^\txp \mathbf{F}^{-1}
=
(\mathbf{F}^\txp - \mathbf{I}) 
(\mathbf{F}^\txp)^{-1} (\mathbf{F}^\txe)^{-1}
=
\left[\mathbf{I} - (\mathbf{F}^\txp)^{-1}\right] (\mathbf{F}^\txe)^{-1}.
\end{equation}
The corresponding jump across the interface is
\begin{equation}
\llbracket{\bm{\upbeta}^\txp}^T\rrbracket
=
\left[\mathbf{I} - (\mathbf{F}^{\txp+})^{-1}\right] (\mathbf{F}^{\txe+})^{-1}
-
\left[\mathbf{I} - (\mathbf{F}^{\txp-})^{-1}\right] (\mathbf{F}^{\txe-})^{-1}
\xrightarrow{\mathbf{F}^{\txe+} = \mathbf{F}^{\txe-}}
-\llbracket \mathbf{F}^{-1} \rrbracket.
\end{equation}
Thus, Eq.~\eqref{alphambpijejoeietcrossdxt} becomes 
\begin{equation}\label{alphajumpFinvncrossdeltaxt}
\bm{\upalpha}
=
\llbracket \mathbf{F}^{-1}\rrbracket 
(\hat{\bm{n}}\times) \delta(x_3).
\end{equation}
This expression shows that the Nye tensor is nonzero only on the interface ($x_3 = 0$​). 
Moreover, $\bm{\upalpha}\hat{\bm{n}} = \mathbf{0}$, meaning that no dislocation line threads through the interface plane.

From Eq.~\eqref{alphajumpFinvncrossdeltaxt}, 
\begin{equation}\label{alphahatejumpFivhatpdeltaxt}
\bm{\upalpha}\hat{\bm{e}}
=
\llbracket \mathbf{F}^{-1}\rrbracket 
\hat{\bm{n}}\times\hat{\bm{e}} \delta(x_3)
=
\llbracket \mathbf{F}^{-1}\rrbracket 
\hat{\bm{p}} \delta(x_3).
\end{equation}
Substituting Eq.~\eqref{alphahatejumpFivhatpdeltaxt} into Eq.~\eqref{bintCHpdXintCbpTFFinvdx}, 
\begin{equation}\label{bpjumpFinvp}
\bm{b}(\bm{p})
= 
\llbracket\mathbf{F}^{-1}\rrbracket 
\hat{\bm{p}} p
=
\llbracket\mathbf{F}^{-1}\rrbracket 
\bm{p},  
\end{equation}
which is exactly the classical FBE (Eq.~\eqref{FBeq}) in the previous section.

\subsection{The generalized Frank-Bilby equation}

The classical FBE predicts only the Burgers vector content needed to accommodate lattice mismatch across an interface. 
However, the most common line defects in interfaces are disconnections, which are characterized not only by a Burgers vector but also by a step height, both of which are defined with respect to the DSC lattice.

To correctly obtain the Burgers vectors of disconnections, we may simply treat the lattices in Fig. ~\ref{FB} as DSC lattices. 
However, this does not provide the step height content of the interface; a second equation is required. 
In this section, we first derive an equation for the total step height intersected by a probe vector on a sharp interface. 
We then couple this result with the classical FBE to obtain a generalized formulation applicable to disconnections.

\subsubsection{Step height density vector and the total step height equation}

\begin{figure}[hbtp]
\includegraphics[width = 0.67\linewidth]{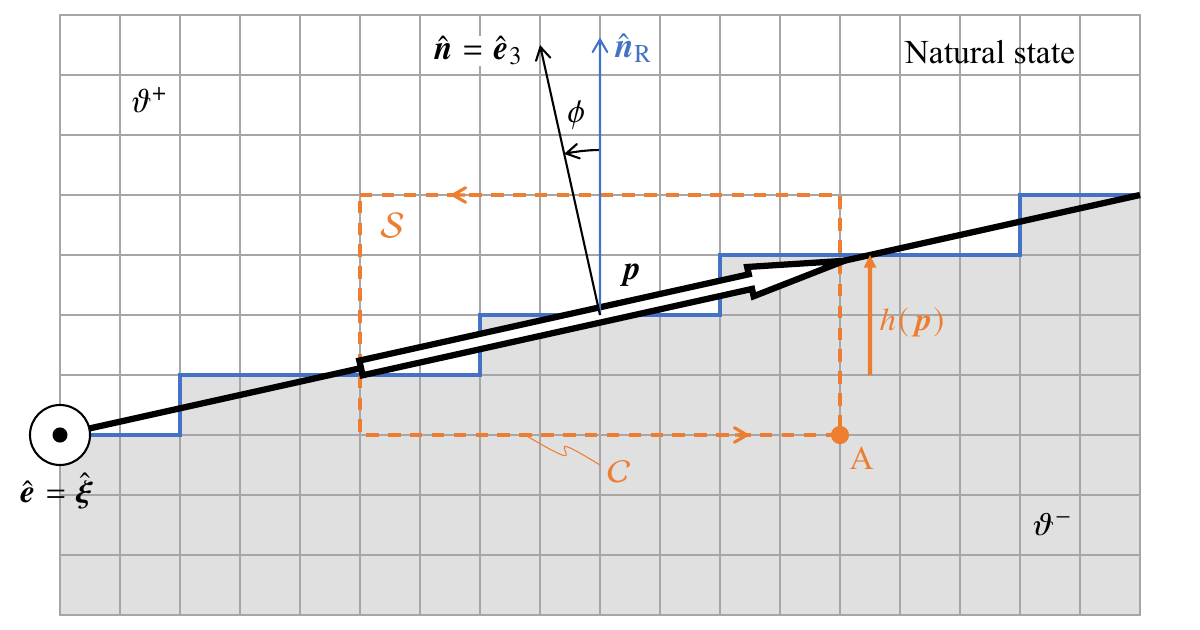}
\caption{\label{stepdensity}
Rectangular circuit $\mathcal{C}$ enclosing a part of interface in NS. 
This interface contains some steps.
The blue lines indicate the microscopically faceted interface morphology; the black line indicate the nominally flat interface.  
$\hat{\bm{n}}$ is the mean interface normal. 
$\hat{\bm{n}}_\txR$ is the normal of the reference interface (zero inclination). 
$\hat{\bm{e}}$ is the normal of the rectangular area. 
Across the interface the order parameter changes from $\vartheta^-$ below to $\vartheta^+$ above. 
} 
\end{figure}

Previously, we showed that the Burgers vector content arises from incompatibility in the deformation gradient $\mathbf{F}$ across an interface. 
By analogy, the step height content arises from the incompatibility of an additional scalar order parameter, denoted by $\vartheta$. 
This parameter takes different constant values inside grains of different orientations; thus, a sharp interface exhibits a jump $\llbracket \vartheta \rrbracket$, as depicted in Fig.~ \ref{stepdensity}. 

A faceted interface contains steps, and the sum of step heights intersected by a probe vector $\bm{p}$ can be obtained by integrating $\vartheta \hat{\bm{n}}_\txR$​ along a Burgers-type circuit $\mathcal{C}$ that encloses $\bm{p}$ (orange dashed line in Fig.~\ref{stepdensity}):
\begin{equation}\label{hCbretabrinvointCetanRdx}
h(\mathcal{C}) = 
-\llbracket\vartheta\rrbracket^{-1}
\oint_{\mathcal{C}} \vartheta \hat{\bm{n}}_\txR 
\cdot\rmd \bm{x}
\xrightarrow{\text{Stokes' theorem}}
\int_{\mathcal{S}} 
\ucurl\left(-\frac{\vartheta \hat{\bm{n}}_\txR}{\llbracket\vartheta\rrbracket}\right) 
\cdot \hat{\bm{e}}\rmd a
\equiv
\int_{\mathcal{S}} \bm{\rho} \cdot \hat{\bm{e}} \rmd a, 
\end{equation}
where $\hat{\bm{n}}_\txR$ is the unit normal vector of the reference interface, which typically corresponds to a facet with zero inclination, and $\hat{\bm{e}}$ is the outward normal of the area $\mathcal{S}$. 
This equation is the direct analog of Eq.~\eqref{bintCHpdXintCbpTFFinvdx}. 
To understand Eq.~\eqref{hCbretabrinvointCetanRdx}, we use the circuit shown in Fig.~\ref{stepdensity} as an example. 
Traverse the circuit counterclockwise from point A to itself (orange dashed line). 
The contributions along the four edges give: 
$\oint_{\mathcal{C}} \vartheta \hat{\bm{n}}_\txR \cdot \rmd\bm{x} = 3a\vartheta^- + a\vartheta^+ - 3a\vartheta^+ - a\vartheta^- = -2a(\vartheta^+ - \vartheta^-) = -2a\llbracket\vartheta\rrbracket$, where $a$ is the lattice parameter; thus, $h(\mathcal{C}) = 2a$. 
Indeed, we see that the total step height along $\bm{p}$ is $2\ell$. 
This validates the step height integral Eq.~\eqref{hCbretabrinvointCetanRdx}. 

From the last equation of Eq.~\eqref{hCbretabrinvointCetanRdx}, we also learn that the step height density vector is defined as 
\begin{equation}\label{stepheightdensityvectordef}
\bm{\rho}
= \ucurl\left(-\frac{\vartheta \hat{\bm{n}}_\txR}{\llbracket\vartheta\rrbracket}\right).
\end{equation}
We will see the explicit meaning of this definition later. 
It is interesting to note that since $\vartheta = \vartheta(\bm{x})$, $\bm{\rho} = \bm{\rho}(\bm{x})$ is defined everywhere in the system, not just on the interface. 
In grain interiors, where $\vartheta$ is constant, $\bm{\rho} = \bm{0}$. 
Thus, although $\bm{\rho}$ is defined in the whole domain, it is nonzero only at the interface. 
This property makes $\bm{\rho}$ naturally compatible with phase-field formulations in which order parameters vary continuously. 

Next, we specialize to the planar interface configuration. 
We compute the curl: 
\begin{equation}
\ucurl(\vartheta\hat{\bm{n}}_\txR)
= e_{ijk} (\vartheta\hat{\bm{n}}_\txR)_{k,j} \hat{\bm{e}}_i
= e_{ijk} \vartheta_{,j}(\hat{\bm{n}}_\txR)_k \hat{\bm{e}}_i. 
\end{equation}
In the configuration of Fig. ~\ref{stepdensity}, the order parameter varies only along $x_3$​; so, $\vartheta_{,j} \ne 0$ only when $j = 3$.
Hence, 
\begin{equation}\label{curletanReetagradnRkei}
\ucurl(\vartheta\hat{\bm{n}}_\txR)
= e_{i3k} \vartheta_{,3}(\hat{\bm{n}}_\txR)_k \hat{\bm{e}}_i
=
\vartheta_{,3} 
\left[-(\hat{\bm{n}}_\txR)_2 \hat{\bm{e}}_1
+ (\hat{\bm{n}}_\txR)_1 \hat{\bm{e}}_2\right]
=
-\vartheta_{,3} \hat{\bm{n}}_\txR \times \hat{\bm{e}}_3
=
-\vartheta_{,3} \hat{\bm{n}}_\txR \times \hat{\bm{n}}. 
\end{equation}
Because the interface is a sharp jump in $\vartheta$,
\begin{equation}
\vartheta(\bm{x}) = \llbracket \vartheta \rrbracket \Theta(x_3) + \vartheta^-
\quad\Rightarrow\quad
\vartheta_{,3} = \llbracket \vartheta \rrbracket \delta(x_3). 
\end{equation}
Equation~\eqref{curletanReetagradnRkei} becomes 
\begin{equation}
\ucurl(\vartheta\hat{\bm{n}}_\txR)
= 
-\llbracket \vartheta \rrbracket (\hat{\bm{n}}_\txR \times \hat{\bm{n}})
\delta(x_3). 
\end{equation}
From Eq.~\eqref{stepheightdensityvectordef}, the step height density vector can be expressed as 
\begin{equation}\label{rhonRtimesndxth}
\bm{\rho}
= 
(\hat{\bm{n}}_\txR \times \hat{\bm{n}})
\delta(x_3),  
\end{equation}
where $\hat{\bm{n}} = \ugrad\vartheta / |\ugrad\vartheta|$. 

For the special case shown in Fig.~\ref{stepdensity}, $\bm{\rho} = \sin\phi \, \hat{\bm{e}} \delta(x_3)$, where $\phi$ is the inclination angle (i.e., the angle between $\hat{\bm{n}}_\txR$ and $\hat{\bm{n}}$). 
The magnitude of $\bm{\rho}$ is $\sin\phi$. 
When $\phi = 0$, the step height density is zero; when $\phi = 90^\circ$, the step height density is $h/p = p/p = 1$.  
The direction of $\bm{\rho}$ is $\hat{\bm{n}}_\txR \times \hat{\bm{n}} = \hat{\bm{e}}$ (but in general, $\hat{\bm{n}}_\txR \times \hat{\bm{n}}$ is not parallel to $\hat{\bm{e}}$). 
We see that for the special case shown in Fig.~\ref{stepdensity}, the step line is indeed parallel to $\hat{\bm{e}}$. 

Using Eq.~\eqref{rhonRtimesndxth}, we evaluate $\bm{\rho} \cdot \hat{\bm{e}}$:
\begin{equation}\label{rhodothatesinphideltaxth}
\bm{\rho} \cdot \hat{\bm{e}}
=
(\hat{\bm{n}}_\txR \times\hat{\bm{n}} \cdot \hat{\bm{e}})
\delta(x_3)
=
(\hat{\bm{n}}_\txR \cdot \hat{\bm{n}} \times \hat{\bm{e}})
\delta(x_3)
=
(\hat{\bm{n}}_\txR \cdot \hat{\bm{p}})
\delta(x_3). 
\end{equation}
Substituting Eq.~\eqref{rhodothatesinphideltaxth} into Eq.~\eqref{hCbretabrinvointCetanRdx}, we obtain
\begin{equation}\label{hpnRdotp}
h(\bm{p}) 
= \hat{\bm{n}}_\txR \cdot \bm{p}. 
\end{equation}
This is the Frank-Bilby-type equation for step height, analogous to Eq.~\eqref{bpjumpFinvp} for Burgers vector. 
Equation~\eqref{hpnRdotp} provides the geometric constraint for the step content of an interface, completing the necessary information to describe disconnections.

\subsubsection{The generalized Frank-Bilby equation}

A general interface containing disconnections should possess both Burgers vector content and step height content. 
To describe the disconnection content in a general interface, we simply combine Eq.~\eqref{bpjumpFinvp} and Eq.~\eqref{hpnRdotp}.
The total ``Burgers-step'' (Burgers vector and step height) content intersected by any probe vector $\bm{p}$ satisfies
\begin{equation}\label{GFB}
\underline{\bm{b}}(\bm{p})
=
\underline{\mathbf{T}} \bm{p}
\quad  \forall \bm{p} \perp \hat{\bm{n}},
\end{equation}
where the extended Burgers-step vector and the geometrical incompatibility are defined as
\begin{equation}
\underline{\bm{b}}
\equiv
\left(\begin{array}{c}
\bm{b} \\ h
\end{array}\right), 
\quad
\underline{\mathbf{T}}
\equiv
\left(\begin{array}{c}
\llbracket\mathbf{F}^{-1}\rrbracket  \\
\hat{\bm{n}}_\txR^T
\end{array}\right). 
\end{equation}
This is the \emph{generalized Frank-Bilby equation (GFBE)}. 
The underline labels an extended quantity, and $\underline{\bm{b}}$ is termed the \emph{Burgers-step vector} of a disconnection. 

When an interface contains only a single mode of disconnections, with Burgers vector $\bm{b}$ and step height $h$, the Nye tensor (Burgers vector density tensor) $\bm{\upalpha}$ and the step height density vector $\bm{\rho}$ must kinematically coupled. 
For an arbitrary direction lying in the interface $\hat{\bm{e}}$, $\bm{\rho} \cdot \hat{\bm{e}}$ represents the density of the step height associated with the disconnection lines parallel to $\hat{\bm{e}}$. 
Multiplying this step density by $\bm{b}/h$ converts step height density into Burgers vector density.
Thus the Burgers vector density for lines parallel to $\hat{\bm{e}}$ is
\begin{equation}
\bm{\upalpha} \hat{\bm{e}}
=
(\bm{\rho} \cdot \hat{\bm{e}}) 
\frac{\bm{b}}{h}
=
\left(\frac{\bm{b}}{h} \otimes \bm{\rho}\right)
\hat{\bm{e}},
\quad \forall \hat{\bm{e}} \perp \hat{\bm{n}}
\quad\Rightarrow\quad
\bm{\upalpha}
=
\frac{\bm{b}}{h} \otimes \bm{\rho}
=
\beta \hat{\bm{b}} \otimes \bm{\rho},  
\end{equation}
where $\beta \equiv |\bm{b}|/h$ is the shear-coupling factor. 
This coupling law applies only when a single disconnection mode is present. 
When multiple disconnection modes coexist, there is no simple linear relationship between $\bm{\upalpha}$ and $\bm{\rho}$.

\subsection{The quantized generalized Frank-Bilby equation}
\label{sec:qGFBE}

The structure of an interface in a crystalline material can be viewed as composed by coherent terraces partitioned by a disconnection network. 
We adopt a bicrystal containing a coherent interface, constructed by appropriately rotating, stretching, and/or shearing the two crystals, as the reference state. 
The bicrystallography of the reference state determines a discrete set of admissible disconnection modes $\{\underline{\bm{b}}^{(m)} = (\bm{b}^{(m)}, h^{(m)})\}$, where $m$ ($=1, \cdots, M$) is the mode index. 
An eﬀiicient numerical algorithm for enumerating all possible disconnection modes has been developed by Admal et al.~\cite{admal2022interface}. 
The central problem addressed here is the following: given the deformation gradients $\mathbf{F}^+$ and $\mathbf{F}^-$ that map the crystals in the reference state to those in the natural state, the unit normal vector of the reference interface plane $\hat{\bm{n}}_\text{R}$, and a list of disconnection modes $\{\underline{\bm{b}}^{(m)}\}$ ($m=1, \cdots, M$), determine the disconnection line directions $\{\hat{\bm{\xi}}^{(m)}\}$ and their spacings $\{d^{(m)}\}$.

Define the reciprocal spacing vector: 
\begin{equation}
\bm{\eta}^{(m)}
=
\frac{\hat{\bm{n}} \times \hat{\bm{\xi}}^{(m)}}{d^{(m)}}. 
\end{equation}
The direction of $\bm{\eta}^{(m)}$ is parallel to the interface plane and perpendicular to the disconnection line; its magnitude equals to the reciprocal disconnection line spacing. 
So, the problem becomes how to determine $\{\hat{\bm{\xi}}^{(m)}\}$.
The number of the $m^\text{th}$ set of disconnection lines intersecting $\bm{p}$ is $\bm{\eta}^{(m)} \cdot \bm{p}$.
Hence, the total Burgers-step content is 
\begin{equation}\label{Stbpsummgmdptbm}
\underline{\bm{b}}(\bm{p})
= 
\sum_{m=1}^M 
\left(\bm{\eta}^{(m)} \cdot \bm{p}\right)
\underline{\bm{b}}^{(m)}
= 
\sum_{m=1}^M
\left(\underline{\bm{b}}^{(m)} \otimes \bm{\eta}^{(m)}\right) \bm{p}
\quad
\forall \bm{p} \perp \hat{\bm{n}}. 
\end{equation}
Combining Eq.~\eqref{Stbpsummgmdptbm} and Eq.~\eqref{GFB} yields:
\begin{equation}\label{summubmotetamuKpzero}
\left[
\sum_{m=1}^M
\left(\underline{\bm{b}}^{(m)} \otimes \bm{\eta}^{(m)}\right)
-
\underline{\mathbf{T}}
\right] 
\bm{p}
=
\bm{0}
\quad
\forall \bm{p} \perp \hat{\bm{n}}.
\end{equation}
Let $\check{\bm{p}}$ denote an arbitrary vector in a three-dimensional space. 
Then, any vector within the interface plane can be expressed as $\bm{p} = (\mathbf{I} - \hat{\bm{n}} \otimes \hat{\bm{n}}) \check{\bm{p}}$.  
Thus, Eq.~\eqref{summubmotetamuKpzero} can be rewritten as 
\begin{equation}\label{summubmotemuKIhnothncp}
\left[
\sum_{m=1}^M
\left(\underline{\bm{b}}^{(m)} \otimes \bm{\eta}^{(m)}\right)
-
\underline{\mathbf{T}}
\right] 
(\mathbf{I} - \hat{\bm{n}} \otimes \hat{\bm{n}}) \check{\bm{p}}
=
\left[
\sum_{m=1}^M
\left(\underline{\bm{b}}^{(m)} \otimes \bm{\eta}^{(m)}\right)
-
\underline{\mathbf{T}}^\parallel
\right] 
\check{\bm{p}}
=
\mathbf{0}
\quad
\forall \check{\bm{p}} \in \mathbb{R}^3,
\end{equation}
where the interfacial incompatibility tensor is defined as
\begin{equation}\label{uKparauKInotn}
\underline{\mathbf{T}}^\parallel_{4\times 3} 
\equiv 
\underline{\mathbf{T}} (\mathbf{I} - \hat{\bm{n}} \otimes \hat{\bm{n}})
=
\left(\begin{array}{c}
\llbracket \mathbf{F}^{-1} \rrbracket \\
\hat{\bm{n}}_\text{R}^T
\end{array}\right)
(\mathbf{I} - \hat{\bm{n}} \otimes \hat{\bm{n}}).
\end{equation}
To ensure that Eq.~\eqref{summubmotemuKIhnothncp} holds for any $\check{\bm{p}} \in \mathbb{R}^3$, we must have
\begin{equation}\label{sumkBNTKpara}
\sum_{m=1}^M
\left(\underline{\bm{b}}^{(m)} \otimes \bm{\eta}^{(m)}\right)
=
\underline{\mathbf{T}}^\parallel
\quad\Rightarrow\quad
\underline{\mathbf{B}} \mathbf{N}^T 
=
\underline{\mathbf{T}}^\parallel, 
\end{equation}
where
\begin{equation}
\underline{\mathbf{B}}_{4\times M}
\equiv
\left(\begin{array}{ccc}
\underline{\bm{b}}^{(1)} &
\cdots &
\underline{\bm{b}}^{(M)}
\end{array}\right)
\text{ and }
\mathbf{N}_{3\times M}
\equiv
\left(\begin{array}{ccc}
\bm{\eta}^{(1)} &
\cdots &
\bm{\eta}^{(M)}
\end{array}\right)
\end{equation}
Equation~\eqref{sumkBNTKpara} is the quantized FB equation, which determines the necessary disconnection network for interface compatibility. 
We can show that the solution $\bm{\eta}^{(m)}$ must lie in the interface plane, i.e., $\bm{\eta}^{(m)} \cdot \hat{\bm{n}} = 0$. 
Left multiplying Eq.~\eqref{sumkBNTKpara} by $\hat{\bm{n}}$, 
\begin{equation}
\sum_{m=1}^{M}
(\bm{\eta}^{(m)} \cdot \hat{\bm{n}}) \underline{\bm{b}}^{(m)}
=
\underline{\mathbf{T}}^\parallel \hat{\bm{n}}
=
\underline{\mathbf{T}}(\mathbf{I} - \hat{\bm{n}} \otimes \hat{\bm{n}}) \hat{\bm{n}}
=
\mathbf{0}_{4\times 1}.
\end{equation}
Because $\underline{\bm{b}}^{(1)}, \cdots, \underline{\bm{b}}^{(M)}$ are linearly independent, all coefficients must be zero, i.e., $\bm{\eta}^{(m)} \cdot \hat{\bm{n}} = 0$ ($m = 1, \cdots, M$).

Equation~\eqref{sumkBNTKpara} represents a system of linear equations. 
Such a linear system may have no solution, a unique solution, or infinitely many solutions. 
The conditions corresponding to each case are summarized follows. 
\begin{itemize}
\item
\emph{No solution:}
Existence of solution is guaranteed by the \emph{condition of consistency}: the columns of $\underline{\mathbf{T}}^\parallel$ lie in the span of the columns of $\underline{\mathbf{B}}$ (i.e., each column of $\underline{\mathbf{T}}^\parallel$ is a linear combination of $\{\underline{\bm{b}}^{(k)}\}$).
This condition can be expressed as ``$\text{Col}(\underline{\mathbf{T}}^\parallel) \subseteq \text{Col}(\underline{\mathbf{B}})$''. 
In practice, we can check whether the augmented matrix $\underline{\mathbf{T}}^\parallel$ introduces no additional rank of $\underline{\mathbf{B}}$: 
\begin{equation}\label{rankBArankB}
\text{rank}([\begin{array}{c | c} \underline{\mathbf{B}} & \underline{\mathbf{T}}^\parallel \end{array}]) 
= \text{rank}(\underline{\mathbf{B}}).
\end{equation} 
If this holds, then $\text{Col}(\underline{\mathbf{T}}^\parallel) \subseteq \text{Col}(\underline{\mathbf{B}})$; otherwise, $\text{Col}(\underline{\mathbf{T}}^\parallel) \nsubseteq \text{Col}(\underline{\mathbf{B}})$.
When the condition of consistency fails, i.e., $\text{Col}(\underline{\mathbf{T}}^\parallel) \nsubseteq \text{Col}(\underline{\mathbf{B}})$, Eq.~\eqref{sumkBNTKpara} has no solution. 
Two strategies can be considered to restore uniqueness of the solution:
(i) introduce additional linearly independent Burgers vectors, or (ii) impose constraints on $\underline{\mathbf{T}}^\parallel$ that make some of its columns linearly dependent. 

\item
\emph{Unique solution:}
When the condition of consistency is satisfied, i.e., $\text{Col}(\underline{\mathbf{T}}^\parallel) \subseteq \text{Col}(\underline{\mathbf{B}})$, Eq.~\eqref{sumkBNTKpara} must have at least one solution.
The uniqueness of solution is guaranteed by the \emph{full-rank condition}: $\text{rank}(\underline{\mathbf{B}}) = M$, i.e., the rank of $\underline{\mathbf{B}}$ is equal to the number of disconnection arrays. 
Combined with Eq.~\eqref{rankBArankB}, the sufficient and necessary condition for uniqueness of solution is
\begin{equation}
\text{rank}([\begin{array}{c | c} \underline{\mathbf{B}} & \underline{\mathbf{T}}^\parallel \end{array}]) 
= \text{rank}(\underline{\mathbf{B}})
= M.
\end{equation} 
The unique solution is obtained by 
\begin{equation}\label{NATBTplus}
\mathbf{N}
=
(\underline{\mathbf{T}}^\parallel)^T (\underline{\mathbf{B}}^T)^*, 
\end{equation}
where ``$*$'' denotes the Moore-Penrose pseudoinverse.
In particular, when $\text{rank}(\underline{\mathbf{B}}) = M = 4$, the condition $\text{Col}(\underline{\mathbf{T}}^\parallel) \subseteq \text{Col}(\underline{\mathbf{B}})$ automatically holds. 
The formula Eq.~\eqref{NATBTplus} reduces to 
\begin{equation}\label{NATBinvT}
\mathbf{N}
=
(\underline{\mathbf{T}}^\parallel)^T \underline{\mathbf{B}}^{-T}.
\end{equation}
(Obviously, in this case $\underline{\mathbf{B}}^T$ is invertible.)

\item
\emph{Infinitely many solutions:}
The sufficient and necessary condition for the existence of infinitely many solutions is $\text{Col}(\underline{\mathbf{T}}^\parallel) \subseteq \text{Col}(\underline{\mathbf{B}})$ and $\text{rank}(\underline{\mathbf{B}}) < M$.
Additional physical criteria, such as minimizing the total disconnection energy, are needed to select a physically meaningful result. 
\end{itemize}
Table~\ref{tab:solutionsummarydisc} categorizes all possible cases and guides practical implementation.  
Because checking the consistency condition directly can be expensive, the table also provides a practical decision workflow.

\begin{table}
\caption{\label{tab:solutionsummarydisc}
Solution summary for $\underline{\mathbf{B}}\mathbf{N}^T = \underline{\mathbf{T}}^\parallel$, where $\mathbf{N}$ is the unknown. 
$\underline{\mathbf{B}}$ is $4\times M$, $\mathbf{N}$ is $3\times M$, and $\underline{\mathbf{T}}^\parallel$ is $4\times 3$.
}
\begin{tabular*}{\columnwidth}{@{\extracolsep{\fill}} c c c c c c @{}}
\toprule
\multicolumn{3}{c}{Conditions} & Types of solutions & Unique solution & Case \\
\midrule
\multirow{3}{*}{$M = 4$} & \multicolumn{2}{c}{$\text{rank}(\underline{\mathbf{B}}) = 4$} & Unique solution & $\mathbf{N} = (\underline{\mathbf{T}}^\parallel)^T \underline{\mathbf{B}}^{-T}$ & (1) \\
\cmidrule{2-6}
 & \multirow{2}{*}{$\text{rank}(\underline{\mathbf{B}}) < 4$} & $\text{Col}(\underline{\mathbf{T}}^\parallel) \subseteq \text{Col}(\underline{\mathbf{B}})$ & Infinite solutions & -- & (2) \\
\cmidrule{3-6}
 & & $\text{Col}(\underline{\mathbf{T}}^\parallel) \nsubseteq \text{Col}(\underline{\mathbf{B}})$ & No solution & -- & (3) \\
\midrule
\multirow{3}{*}{$M < 4$} & \multicolumn{2}{c}{$\text{Col}(\underline{\mathbf{T}}^\parallel) \nsubseteq \text{Col}(\underline{\mathbf{B}})$} & No solution & -- & (4) \\
\cmidrule{2-6}
 & \multirow{2}{*}{$\text{Col}(\underline{\mathbf{T}}^\parallel) \subseteq \text{Col}(\underline{\mathbf{B}})$} & $\text{rank}(\underline{\mathbf{B}}) = M$ & Unique solution & $\mathbf{N} = (\underline{\mathbf{T}}^\parallel)^T (\underline{\mathbf{B}}^T)^*$ & (5) \\
\cmidrule{3-6}
 & & $\text{rank}(\underline{\mathbf{B}}) < M$ & Infinite solution & -- & (6) \\
\midrule
\multirow{2}{*}{$M > 4$} & \multicolumn{2}{c}{$\text{Col}(\underline{\mathbf{T}}^\parallel) \subseteq \text{Col}(\underline{\mathbf{B}})$} & Infinite solutions & -- & (7) \\
\cmidrule{2-6}
 & \multicolumn{2}{c}{$\text{Col}(\underline{\mathbf{T}}^\parallel) \nsubseteq \text{Col}(\underline{\mathbf{B}})$} & No solution & -- & (8) \\
 \bottomrule
\end{tabular*}
\end{table}

\section{Misoriented coherent twin boundaries}
\label{sec:misoriented_CTB}

\subsection{Grain boundaries with arbitrary geometrical parameters}

A grain boundary (GB) is a special type of interface in crystalline materials across which the two adjoining phases have identical crystal structure and lattice parameter, but differ in crystallographic orientation. 
Thus, unlike a general heterophase interface, the two phases separated by a GB do not require stretch or shear transformations to map their RS to their NS. 
Instead, each phase can be generated from a common reference lattice by a rigid-body rotation. 
Consequently, the deformation gradients of the two grains reduce to rotation tensors: $\mathbf{F}^+ = \mathbf{R}^+$ and $\mathbf{F}^- = \mathbf{R}^-$, where $\mathbf{R}$ denotes a rotation tensor. 
To specialize the qGFBE (Eq.~\eqref{sumkBNTKpara}) for the study of GBs, we replace $\llbracket \mathbf{F}^{-1} \rrbracket$ with $\llbracket \mathbf{R}^{-1} \rrbracket$. 
For the ``$+$'' grain, suppose that its orientation relative to the chosen reference state (RS) is obtained by a rotation through an angle $\theta^+$ about a unit rotation axis $\hat{\bm{o}}^+$. 
The corresponding rotation tensor is given by Rodrigues' formula,
\begin{equation}\label{RpluscosthetaIcosthetaootostot}
\mathbf{R}^+
= \cos\theta^+ \mathbf{I}
+ (1 - \cos\theta^+) (\hat{\bm{o}}^+\otimes\hat{\bm{o}}^+)
+ \sin\theta^+(\hat{\bm{o}}^+\times). 
\end{equation}
The inverse of $\mathbf{R}^+$ is obtained simply by reversing the sign of the rotation angle: 
\begin{equation}
(\mathbf{R}^+)^{-1}
= \cos\theta^+ \mathbf{I}
+ (1 - \cos\theta^+) (\hat{\bm{o}}^+\otimes\hat{\bm{o}}^+)
- \sin\theta^+(\hat{\bm{o}}^+\times). 
\end{equation}
Similarly, the inverse rotation tensor for the ``$-$'' grain is 
\begin{equation}
(\mathbf{R}^-)^{-1}
= \cos\theta^- \mathbf{I}
+ (1 - \cos\theta^-) (\hat{\bm{o}}^-\otimes\hat{\bm{o}}^-)
- \sin\theta^-(\hat{\bm{o}}^-\times). 
\end{equation}
The orientation relationship between two grains is fully specified by a single relative rotation, namely a misorientation axis and a misorientation angle. 
Thus, one possible choice of RS is to take one of the grains, say the ``$-$'' grain, as the reference lattice and rotate the other grain relative to it. 
This choice is simple, but it treats the two grains asymmetrically.
For GBs, a more convenient and symmetric choice is the \emph{median lattice}. 
The median lattice is defined such that the two grains are generated from it by equal and opposite rotations about a common axis $\hat{\bm{o}}$. 
Specifically, if the total misorientation angle between the two grains is $\theta$, then $\hat{\bm{o}} = \hat{\bm{o}}^+ = \hat{\bm{o}}^-$ and $\theta^+ = -\theta^- = \theta/2$, where $\theta$ is the misorientation angle. 
With the median lattice chosen as RS, 
\begin{equation}\label{brRinvbrmtsinttocp}
\llbracket \mathbf{R}^{-1} \rrbracket
=
- 2\sin(\theta/2) (\hat{\bm{o}}\times).
\end{equation}
The quantity entering the qGFBE is the interfacial incompatibility tensor. According to Eq.~\eqref{uKparauKInotn}, this is
\begin{equation}\label{Kpara2sintheta2otimesInotn}
\underline{\mathbf{T}}^\parallel
=
\left(\begin{array}{c}
-2\sin(\theta/2) (\hat{\bm{o}}\times) \\
\hat{\bm{n}}_\txR^T
\end{array}\right)
(\mathbf{I} - \hat{\bm{n}}\otimes\hat{\bm{n}}),
\end{equation}
where $\hat{\bm{n}}_\txR$ is the interface normal in the RS (median lattice) and $\hat{\bm{n}}$ is the interface normal in the NS. 

We next consider the important special case of a symmetric tilt GB. For a symmetric tilt boundary, the misorientation axis lies in the GB plane and is perpendicular to the GB normal. 
Hence, $\hat{\bm{o}} \perp \hat{\bm{n}}$. 
Furthermore, under the median-lattice construction, the interface plane is not tilted with respect to the RS, so that $\hat{\bm{n}} = \hat{\bm{n}}_\txR$. 
We can then introduce an orthonormal coordinate system adapted to the boundary geometry: $\hat{\bm{e}}_1 = \hat{\bm{o}}$, $\hat{\bm{e}}_2 = \hat{\bm{n}} \times \hat{\bm{o}}$, $\hat{\bm{e}}_3 = \hat{\bm{n}}$.
Then, Eq.~\eqref{Kpara2sintheta2otimesInotn} simplifies to
\begin{equation}\label{uKparasintmat}
\underline{\mathbf{T}}^\parallel
=
-2\sin(\theta/2)
\left(\begin{array}{ccc}
0 & 0 & 0 \\
0 & 0 & 0 \\
0 & 1 & 0 \\
0 & 0 & 0 \\
\end{array}\right).
\end{equation} 
If a disconnection mode, $(0, 0, b, 0)^T$, is crystallographically admissible, then a single set of such disconnections is sufficient to accommodate the misorientation. 
However, the admissible disconnection modes are constrained by the DSC lattice of the boundary. 
If this mode is not contained in the DSC lattice, multiple disconnection modes must combine so that their net Burgers vector and step content satisfy the qGFBE.

\subsection{Symmetric tilt boundaries as deviations from $\Sigma 3$ coherent twin boundaries}

\begin{figure}[tb]
\includegraphics[width=0.9\columnwidth]{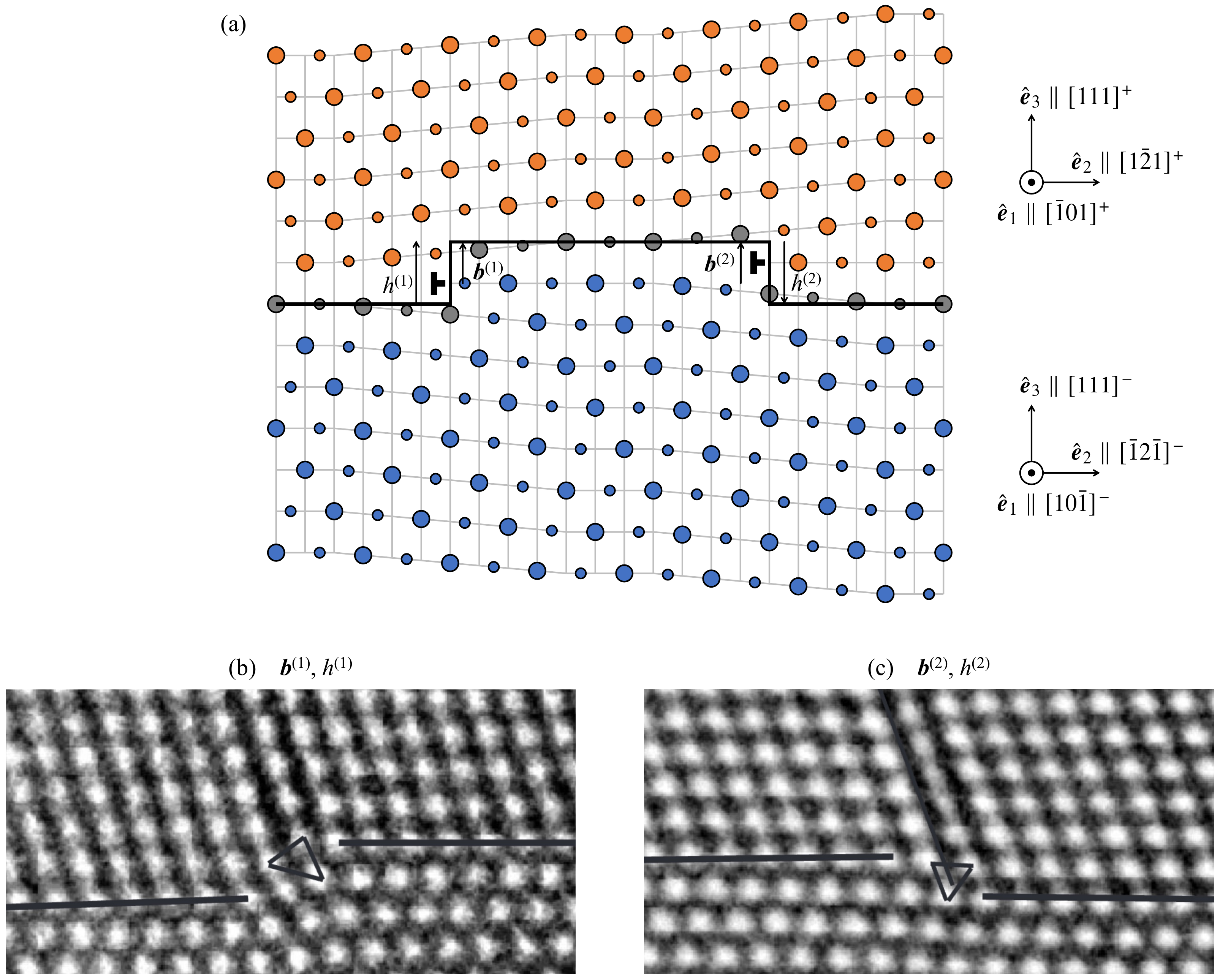}
\caption{\label{NiSigma3} 
Disconnection modes of the $\Sigma 3$ $(111)$ $[10\bar{1}]$ coherent twin boundary.
(a) DSC lattice of the $\Sigma 3$ CTB. 
Blue and orange points represent lattice points belonging to the ``$-$'' and ``$+$'' grains, respectively. 
The point size distinguishes lattice points in different layers parallel to the page. 
Gray points denote coincidence site lattice points along the CTB. 
Light gray grid indicates the DSC lattice, and black line denotes the GB position. 
Two disconnections of different modes are introduced.
(b, c) Atomic structures of the two disconnection modes observed by HRTEM~\cite{marquis2005structural}.} 
\end{figure}

We now apply the above formulation to a misoriented coherent twin boundary (CTB). 
The ideal reference boundary is chosen to be the $\Sigma 3$ CTB in an FCC crystal. 
This ideal CTB has boundary plane $(111)$, and its normal vector in the RS is $\hat{\bm{n}}_\txR = [111]$ (see Fig.~\ref{NiSigma3}). 
We introduce a coordinate system tied to the ideal CTB: $\hat{\bm{e}}_1 \parallel [10\bar{1}]$, $\hat{\bm{e}}_2 \parallel [\bar{1}2\bar{1}]$, $\hat{\bm{e}}_3 \parallel [111]$.
The direction $\hat{\bm{e}}_1$ is chosen as the tilt axis. 
A misoriented CTB is then constructed by rotating the two grains symmetrically about $\hat{\bm{e}}_1$ by equal and opposite angles $\pm \Delta\theta/2$, where $\Delta\theta$ iis the deviation of the misorientation angle from that of the ideal $\Sigma 3$ CTB. 
This construction should be interpreted as follows. The ideal $\Sigma 3$ CTB is used as the RS. The misoriented CTB is the NS generated by applying a small additional symmetric tilt to the two grains. Because the additional rotation is symmetric about an in-plane axis, the boundary plane remains untilted relative to the RS. 
Hence, $\hat{\bm{n}} = \hat{\bm{n}}_\txR = \hat{\bm{e}}_3$. 
The geometry is therefore identical to the symmetric tilt case discussed above, with $\hat{\bm{o}} \to \hat{\bm{e}}_1$ and $\theta \to \Delta\theta$.
Accordingly, Eq.~\eqref{uKparasintmat} becomes
\begin{equation}
\underline{\mathbf{T}}^\parallel
=
-2\sin(\Delta\theta/2)
\left(\begin{array}{ccc}
0 & 0 & 0 \\
0 & 0 & 0 \\
0 & 1 & 0 \\
0 & 0 & 0 \\
\end{array}\right).
\end{equation}
Thus, the deviation from the ideal CTB introduces a tangential mismatch that is formally equivalent to that of a symmetric tilt GB. The difference from a general GB is that the admissible defect content is not arbitrary; it is constrained by the DSC lattice of the $\Sigma 3$ CTB.

The relevant disconnection modes can be identified either geometrically from the DSC lattice, as illustrated in Fig.~\ref{NiSigma3}, or analytically using the Smith-normal-form crystallographic method developed by Admal et al.~\cite{admal2022interface}. 
For the $\Sigma 3$ $(111)$ CTB considered here, two elementary disconnection modes are sufficient. Their Burgers vectors and step heights are collected in the matrix:
\begin{equation}
\underline{\mathbf{B}}
=
\left(\begin{array}{cc}
\underline{\bm{b}}^{(1)} & \underline{\bm{b}}^{(2)}
\end{array}\right)
=
a
\left(\begin{array}{cc}
0 & 0 \\
0 & 0 \\
\frac{\sqrt{3}}{3} & \frac{\sqrt{3}}{3} \\
\frac{\sqrt{3}}{2} & -\frac{\sqrt{3}}{2} \\
\end{array}\right),
\end{equation} 
where $a$ is the lattice parameter. Each column represents one disconnection mode. 
The two modes have identical Burgers vectors but opposite step heights.

For the misoriented CTB, $\text{rank}([\begin{array}{c|c} \underline{\mathbf{B}} & \underline{\mathbf{T}}^\parallel \end{array}]) = \text{rank}( \underline{\mathbf{B}}) = 2 = M$.
According to Case (5) in Table~\ref{tab:solutionsummarydisc}, the solution of the qGFBE is unique. 
The unique solution can be obtained as
\begin{equation}
\mathbf{N}
=
(\underline{\mathbf{T}}^\parallel)^T
(\underline{\mathbf{B}}^T)^*
=
-\frac{\sqrt{3}}{a} \sin(\Delta\theta/2)
\left(\begin{array}{cc}
0 & 0 \\
1 & 1 \\
0 & 0
\end{array}\right)
=
\left(\begin{array}{cc}
\bm{\eta}^{(1)} & \bm{\eta}^{(2)}
\end{array}\right).
\end{equation} 
Therefore, the line directions of the two sets of disconnections are $\hat{\bm{\eta}}^{(1)} \times \hat{\bm{n}} = \hat{\bm{\eta}}^{(2)} \times \hat{\bm{n}} = -\hat{\bm{e}}_1$.
The line spacings of the two sets are $d^{(1)} = d^{(2)} = 1/|\bm{\eta}^{(1)}| = a/(\sqrt{3} \sin(\Delta\theta/2))$. 
A mode-(1) disconnection and a mode-(2) disconnection have opposite step heights and therefore form a step-free pair. 
The net Burgers vector of such a pair is $b = 2a/\sqrt{3}$. 
Since the two modes have the same density and the same spacing, the paired disconnections may be regarded as an array of pure edge dislocations with spacing $d = d^{(1)} = d^{(2)} = b/(2\sin(\Delta\theta/2))$; this is precisely Frank's formula for the spacing of an array of edge dislocations accommodating a small tilt misorientation.

\section{Twist heterophase interfaces}
\label{sec:twist_heterophase}

In this section, we apply the qGFBE to twist heterophase interfaces. 
The purpose is to show how the same framework can describe disconnection networks in which both lattice mismatch and twist misorientation contribute to the required Burgers-step content. 
We focus on the experimentally studied $(001)$ Au-Pd twist interfaces, for which regular networks of interfacial defects were observed by Hwang et al.~\cite{Hwang1980Interphase}. 
These interfaces provide a useful testing ground because Au and Pd have the same FCC crystal structure but different lattice parameters. 
Thus, unlike a twist GB, the interface contains both an angular mismatch and a lattice mismatch.

Two choices of RS are considered below. 
The first is the simplest possible coincidence construction, in which one Au lattice period is matched to one Pd lattice period in the interface plane. 
We refer to this as the $\Sigma_\text{Au}:\Sigma_\text{Pd}=1:1$ interface. 
The second is a higher-order coincidence construction in which nine Au lattice sites and ten Pd lattice sites are contained in one coincidence cell. 
This case is denoted as the $\Sigma_\text{Au}:\Sigma_\text{Pd}=9:10$ interface. 
The latter construction substantially reduces the in-plane strain required to match the two lattices and is therefore more appropriate for describing Au-\-Pd interfaces with a small residual angular deviation from this special orientation relationship.

\begin{figure}[tb]
\includegraphics[width=0.99\columnwidth]{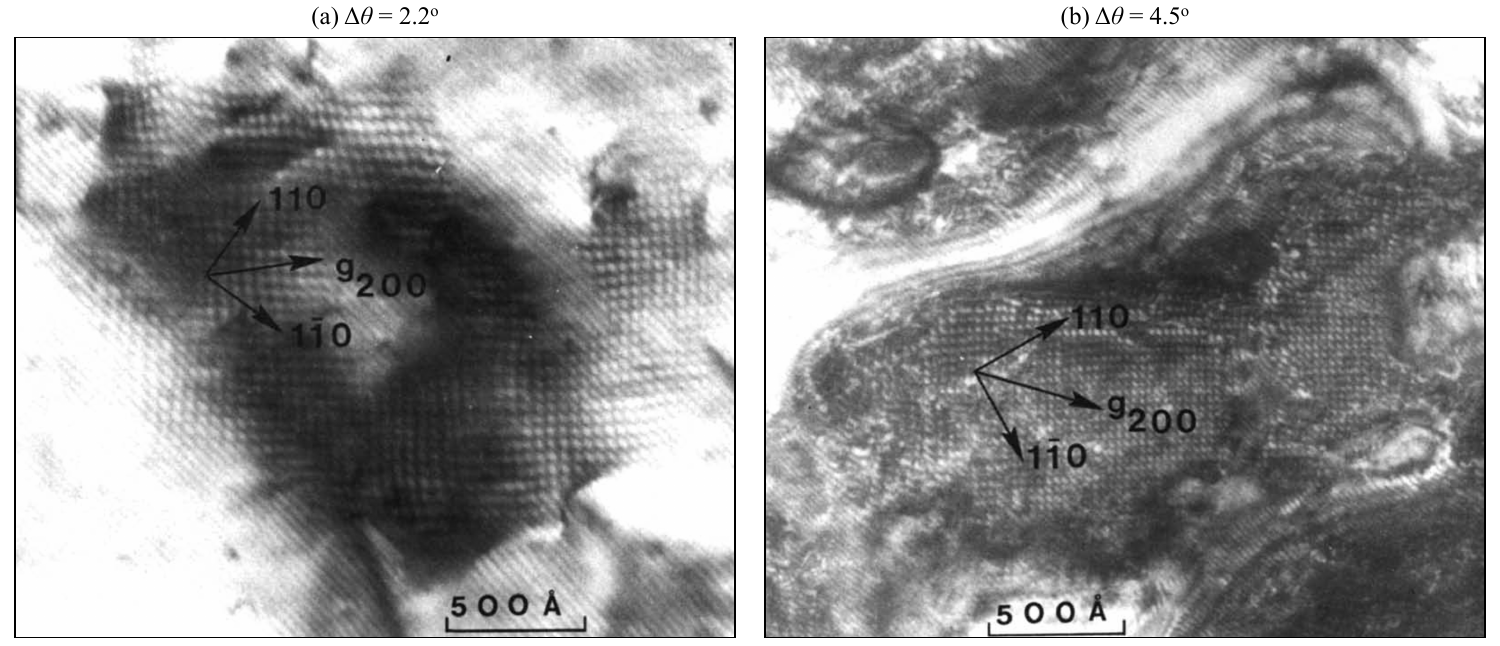}
\caption{\label{fig:exptAuPd}
Disconnection networks experimentally observed in $(001)$ twist interfaces of Au-Pd~\cite{Hwang1980Interphase}.
The disconnection lines are neither parallel nor perpendicular to the $\langle 110 \rangle$ directions, which are the orientations of the Burgers vectors. 
(a) Twist angle $\Delta\theta \sim 2.2^\circ$, with a disconnection spacing of $\sim 47$~{\AA}. 
(b) Twist angle $\Delta\theta \sim 4.5^\circ$, with a disconnection spacing of $\sim 31$~{\AA}. 
} 
\end{figure}

\subsection{The $\Sigma_\text{Au} : \Sigma_\text{Pd} = 1:1$ interface}

We first consider the simplest reference construction for a $(001)$ Au-Pd twist interface. 
Starting from the NS, the Au and Pd crystals are rotated about their common $[001]$ axis by equal and opposite angles, $\pm\Delta\theta/2$, so that their in-plane crystallographic directions $[100]$, $[010]$, and $[001]$ become aligned. 
The two crystals are then homogeneously strained so that their $(001)$ planes match exactly. 
This aligned and coherent configuration is chosen as the RS.

In this RS, if both lattices are extended to fill the whole space, every Au lattice point coincides with a Pd lattice point. 
Hence, the CSL is identical to the common strained lattice. 
For this special construction, the DSC lattice is also identical to the CSL. 
Since one Au lattice period is matched to one Pd lattice period in the interface plane, the reciprocal coincidence-site densities of Au and Pd are both equal to one. 
We therefore denote this interface as the $\sigma_\text{Au} : \sigma_\text{Pd} = 1:1$ interface. 
We introduce an orthonormal coordinate system in the RS: $\hat{\bm{e}}_1 \parallel [1\bar{1}0]$, $\hat{\bm{e}}_2 \parallel [110]$, $\hat{\bm{e}}_3 = \hat{\bm{n}}_\txR \parallel [001]$, where $\hat{\bm{n}}_\text{R}$ is the interface normal in the RS. The interface plane is therefore spanned by $\hat{\bm{e}}_1$ and $\hat{\bm{e}}_2$.

To bring the Au and Pd lattices from the RS back to the NS, we apply homogeneous deformation gradients $\mathbf{F}_\text{Au}$ and $\mathbf{F}_\text{Pd}$ to the two phases. 
We take the interface normal $\hat{\bm{n}}$ to point from Au to Pd. 
With this convention, $\llbracket \mathbf{F}^{-1} \rrbracket = \mathbf{F}_\text{Pd}^{-1} - \mathbf{F}_\text{Au}^{-1}$. 
For Au, the deformation from the RS to the NS consists of an isotropic stretch followed by a rotation about $\hat{\bm{e}}_3$ by an angle $\Delta\theta/2$. 
Thus,
\begin{equation}
\mathbf{U}_\text{Au}
=
\left(\begin{array}{ccc}
r_\text{Au} & 0 & 0 \\
0 & r_\text{Au} & 0 \\
0 & 0 & r_\text{Au}
\end{array}\right),
\quad
\mathbf{R}_\text{Au}
=
\left(\begin{array}{ccc}
\cos(\Delta\theta/2) & \sin(\Delta\theta/2) & 0 \\
-\sin(\Delta\theta/2) & \cos(\Delta\theta/2) & 0 \\
0 & 0 & 1
\end{array}\right),
\quad
\mathbf{F}_\text{Au}
=
\mathbf{R}_\text{Au} \mathbf{U}_\text{Au}. 
\end{equation}
Similarly, for Pd,
\begin{equation}
\mathbf{U}_\text{Pd}
=
\left(\begin{array}{ccc}
r_\text{Pd} & 0 & 0 \\
0 & r_\text{Pd} & 0 \\
0 & 0 & r_\text{Pd}
\end{array}\right),
\quad
\mathbf{R}_\text{Pd}
=
\left(\begin{array}{ccc}
\cos(\Delta\theta/2) & -\sin(\Delta\theta/2) & 0 \\
\sin(\Delta\theta/2) & \cos(\Delta\theta/2) & 0 \\
0 & 0 & 1
\end{array}\right),
\quad
\mathbf{F}_\text{Pd}
=
\mathbf{R}_\text{Pd} \mathbf{U}_\text{Pd}. 
\end{equation}
The stretch ratios are determined by the lattice periodicities of Au and Pd in the RS. 
Let $l_\text{Au}$ and $l_\text{Pd}$ denote the lengths of the shortest CSL vectors along $\hat{\bm{e}}_1$ and $\hat{\bm{e}}_2$, measured with respect to the unstrained Au and Pd lattices, respectively. 
For the present $1:1$ construction, $l_\text{Au} = a_\text{Au}/\sqrt{2}$ and $l_\text{Pd} = a_\text{Pd}/\sqrt{2}$, where $a_\text{Au} = 0.40785$~nm and $a_\text{Pd} = 0.38902$~nm are the lattice parameters for Au and Pd, respectively.
The common RS length is chosen as the arithmetic average: $\bar{l} = (l_\text{Au} + l_\text{Pd})/2$. 
Hence
\begin{equation}
r_\text{Au}
=
\frac{l_\text{Au}}{\bar{l}}
=
\frac{2l_\text{Au}}{l_\text{Au} + l_\text{Pd}},
\quad
r_\text{Pd}
=
\frac{l_\text{Pd}}{\bar{l}}
=
\frac{2l_\text{Pd}}{l_\text{Au} + l_\text{Pd}}.
\end{equation}
Since $a_\text{Au}>a_\text{Pd}$, one has $r_\text{Au} > r_\text{Pd}$. 

The inverse deformation-gradient jump is $\llbracket\mathbf{F}^{-1}\rrbracket = (\mathbf{R}_\text{Pd} \mathbf{U}_\text{Pd})^{-1} - (\mathbf{R}_\text{Au} \mathbf{U}_\text{Au})^{-1}$. 
Because the rotations are about the interface normal, the normal direction is unchanged: $\hat{\bm{n}} = \hat{\bm{n}}_\txR = (0, 0, 1)^T$. 
Therefore, from Eq.~\eqref{uKparauKInotn}, the interfacial incompatibility tensor entering the qGFBE is
\begin{equation}\label{uTpFijhnRiIhnothncssc}
\underline{\mathbf{T}}^\parallel
=
\left(\begin{array}{c}
\llbracket \mathbf{F}^{-1} \rrbracket \\
\hat{\bm{n}}_\txR^{-1}
\end{array}\right)
(\mathbf{I} - \hat{\bm{n}} \otimes \hat{\bm{n}})
=
\left(\begin{array}{ccc}
c & s & 0 \\
-s & c & 0 \\
0 & 0 & 0 \\
0 & 0 & 0
\end{array}\right),
\end{equation}
where
\begin{equation}
c 
\equiv 
\frac{(r_\text{Au} - r_\text{Pd}) \cos(\Delta\theta/2)}{r_\text{Au} r_\text{Pd}}
\text{ and }
s 
\equiv 
\frac{2 \sin(\Delta\theta/2)}{r_\text{Au} r_\text{Pd}}.
\end{equation}
The two scalar quantities $c$ and $s$ have distinct physical origins. 
The quantity $c$ arises from the lattice-parameter mismatch between Au and Pd. 
It remains nonzero even when $\Delta\theta=0$, provided $r_\text{Au}\neq r_\text{Pd}$. 
The quantity $s$, by contrast, arises from the twist misorientation and vanishes when the twist angle is zero. 
Thus, the qGFBE naturally combines the effects of lattice mismatch and twist misorientation in a single interfacial incompatibility tensor.

For the $\Sigma_\text{Au}:\Sigma_\text{Pd}=1:1$ interface, the DSC lattice coincides with the common strained FCC lattice. 
The shortest admissible in-plane Burgers vectors are therefore along $\hat{\bm{e}}_1$ and $\hat{\bm{e}}_2$. 
Since the interface position does not need to change for these in-plane shifts, the corresponding step height is zero. 
The disconnection-mode matrix can be written as
\begin{equation}
\underline{\mathbf{B}}
=
\left(\begin{array}{cc}
\underline{\bm{b}}^{(1)} & \underline{\bm{b}}^{(2)}
\end{array}\right)
=
\frac{\bar{a}}{\sqrt{2}}
\left(\begin{array}{cc}
1 & 0 \\
0 & 1 \\
0 & 0 \\
0 & 0 \\
\end{array}\right),
\end{equation} 
where $\bar{a} = \sqrt{2}\bar{l}$ is the average lattice parameter. 
Because $\text{rank}([\begin{array}{c|c} \underline{\mathbf{B}} & \underline{\mathbf{T}}^\parallel \end{array}]) = \text{rank}( \underline{\mathbf{B}}) = 2 = M$, the problem corresponds to Case (5) in Table~\ref{tab:solutionsummarydisc}. 
Hence, the qGFBE admits a unique solution and the unique solution is
\begin{equation}
\mathbf{N}
=
(\underline{\mathbf{T}}^\parallel)^T
(\underline{\mathbf{B}}^T)^*
=
\frac{\sqrt{2}}{\bar{a}}
\left(\begin{array}{cc}
c & -s \\
s & c \\
0 & 0
\end{array}\right)
=
\left(\begin{array}{cc}
\bm{\eta}^{(1)} & \bm{\eta}^{(2)}
\end{array}\right).
\end{equation} 

For mode (1), the angle between Burgers vector and the line direction is
\begin{align}\label{eq:omega_AuPd_1_1}
\omega^{(1)}
&=
\arccos\left(
\hat{\bm{b}}^{(1)} \cdot (\hat{\bm{\eta}}^{(1)} \times \hat{\bm{n}}_\txR)
\right)
=
\arccos\left(
\frac{s}{\sqrt{c^2 + s^2}}
\right)
=
\arccos\left(
\frac{2\sin(\Delta\theta/2)}{\sqrt{(r_\text{Au} - r_\text{Pd})^2 \cos^2(\Delta\theta/2) + 4\sin^2(\Delta\theta/2)}}
\right)
\nonumber\\
&=
\frac{\pi}{2} - \frac{\Delta\theta}{2}
- \arctan\left(\frac{a_\text{Au} \sin(\Delta\theta)}{a_\text{Pd} - a_\text{Au} \sin(\Delta\theta)}\right).
\end{align}
The disconnection line spacing is
\begin{equation}\label{eq:d_AuPd_1_1_a}
d^{(1)}
=
\frac{1}{\left|\bm{\eta}^{(1)}\right|}
=
\frac{\bar{a} r_\text{Au} r_\text{Pd}}
{\sqrt{2} \sqrt{r_\text{Au}^2 + r_\text{Pd}^2 - 2r_\text{Au} r_\text{Pd} \cos(\Delta\theta)}}
=
\frac{a_\text{Au} a_\text{Pd}}
{\sqrt{2}\sqrt{a_\text{Au}^2 + a_\text{Pd}^2 - 2a_\text{Au} a_\text{Pd} \cos(\Delta\theta)}}.
\end{equation}
Equations~\eqref{eq:omega_AuPd_1_1} and \eqref{eq:d_AuPd_1_1_a} reproduce the results of Jesser and Kuhlmann-Wilsdorf~\cite{Jesser1967Geometry}. 
In the present derivation, however, no skillful geometrical construction of the dislocation network is required. 
Once the RS, the deformation gradients, and the DSC lattice are specified, the orientation and spacing of the disconnection arrays follow directly from the qGFBE.

\subsection{The $\Sigma_\text{Au} : \Sigma_\text{Pd} = 9:10$ interface}

\begin{figure}[tb]
\includegraphics[width=0.73\columnwidth]{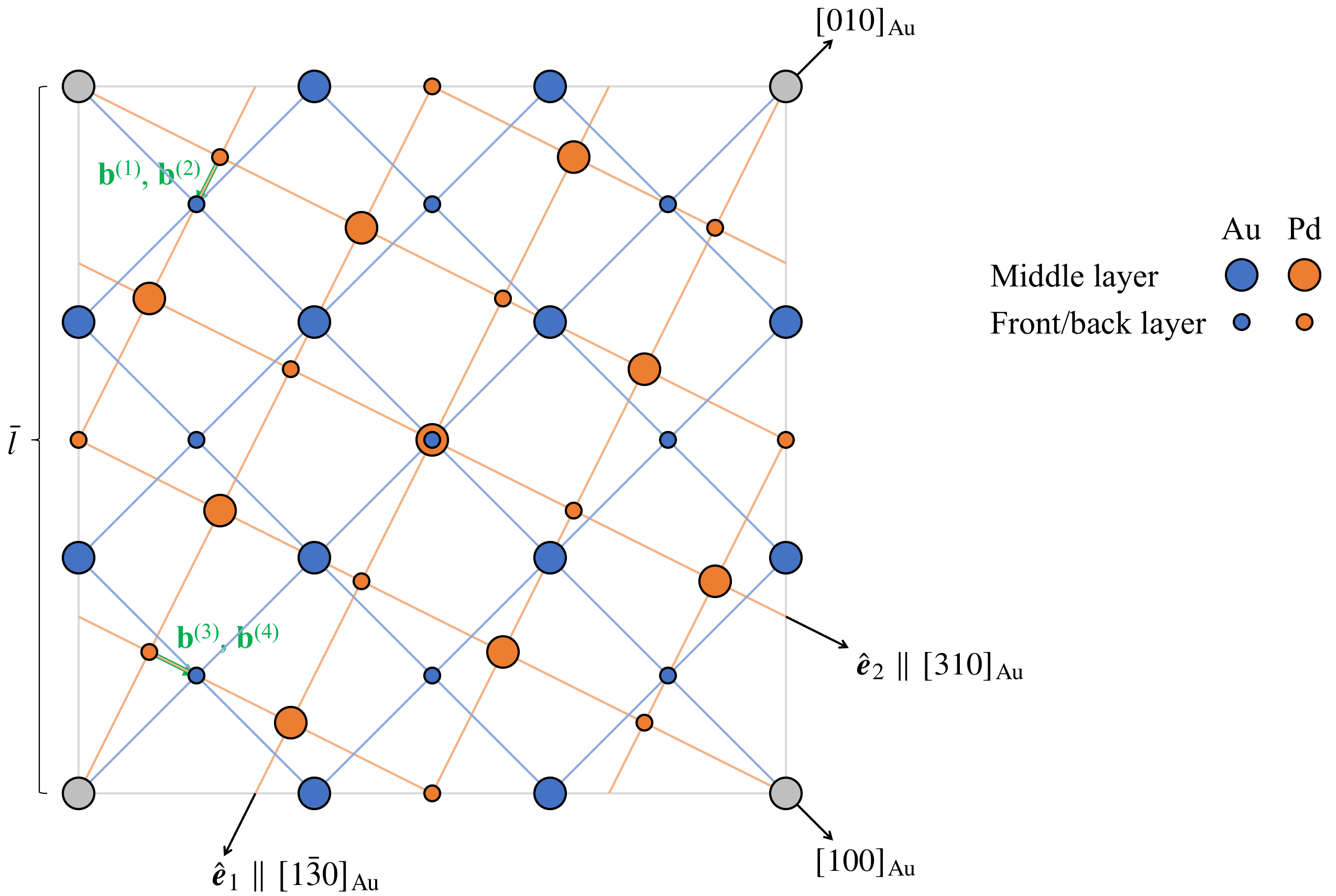}
\caption{\label{fig:Au9Pd10}
Projection along $[001]$ of a CSL cell formed by the strained $(001)$ planes of Au and Pd.
Blue and orange points represent Au and Pd lattice points, respectively. 
Gray points denote coincidence sites. 
The point size indicates different layers parallel to the page.
One CSL cell contains 9 Au lattice points and 10 Pd lattice points; thus, this interface is denoted as $\Sigma_\text{Au}:\Sigma_\text{Pd}=9:10$. 
The two green vectors denote the shortest Burgers vectors of admissible disconnection modes. 
Translating the entire Pd lattice relative to the Au lattice by either green vector moves the coincidence sites to a different plane, indicating that relative translation is coupled to a change of interface position; hence, these modes possess finite step heights.
} 
\end{figure}

We next consider a higher-order coincidence construction for the Au-Pd $(001)$ interface. 
This construction corresponds to a special orientation relationship in which the Au and Pd crystals are rotated relative to one another about $[001]$ by $\theta = 18.43^\circ$.
At this orientation, $[1\bar{1}0]_\text{Au} \parallel [210]_\text{Pd}$, $[110]_\text{Au} \parallel [\bar{1}20]_\text{Pd}$, $[001]_\text{Au} \parallel [001]_\text{Pd}$.
The resulting coincidence cell is shown schematically in Fig.~\ref{fig:Au9Pd10}. 
After this relative rotation, the Au and Pd phases are homogeneously strained so that the three edges of the coincidence cell are exactly aligned. 
This strained coincidence configuration is chosen as the RS.

The in-plane edge lengths of the coincidence cell are: $l_1 = l_2 = (3/\sqrt{2})a_\text{Au} \equiv l_\text{Au} = 0.8652$~nm for Au, and $l_1 = l_2 = \sqrt{5} a_\text{Pd} \equiv l_\text{Pd} = 0.8699$~nm for Pd. 
The normal-period lengths are: $l_3 = a_\text{Au} \equiv l_\text{Au}'$ for Au, and $l_3 = a_\text{Pd} \equiv l_\text{Pd}'$ for Pd.
The common in-plane RS length is $\bar{l} = (l_\text{Au} + l_\text{Pd})/2$, and the in-plane stretch ratios are
\begin{equation}
r_\text{Au}
=
\frac{l_\text{Au}}{\bar{l}},
\quad
r_\text{Pd}
=
\frac{l_\text{Pd}}{\bar{l}}.
\end{equation}
Because $l_\text{Au}<l_\text{Pd}$ for this coincidence construction, $r_\text{Au} < r_\text{Pd}$.
Along the interface normal, we similarly define $\bar{l}' = (l'_\text{Au} + l'_\text{Pd})/2$, and
\begin{equation}
r_\text{Au}'
=
\frac{l_\text{Au}'}{\bar{l}'},
\quad
r_\text{Pd}'
=
\frac{l_\text{Pd}'}{\bar{l}'}.
\end{equation}
Since $a_\text{Au}>a_\text{Pd}$, $r_\text{Au}'>r_\text{Pd}'$.
We introduce a coordinate system associated with the coincidence cell: $\hat{\bm{e}}_1 \parallel [1\bar{3}0]_\text{Au} \parallel [100]_\text{Pd}$, $\hat{\bm{e}}_2 \parallel [310]_\text{Au} \parallel [010]_\text{Pd}$, $\hat{\bm{e}}_3 \parallel [001]_\text{Au} \parallel [001]_\text{Pd}$, as shown in Fig.~\ref{fig:Au9Pd10}. 
The interface normal is $\hat{\bm{n}}_\txR = \hat{\bm{e}}_3$.

A small additional twist deviation $\Delta\theta$ from this special $\Sigma_\text{Au}:\Sigma_\text{Pd}=9:10$ orientation relationship is then introduced by rotating Au and Pd symmetrically about $\hat{\bm{e}}_3$ by $\pm\Delta\theta/2$. 
The interfacial incompatibility tensor has the same form as in the $1:1$ case (Eq.~\eqref{uTpFijhnRiIhnothncssc}), except that the relevant in-plane stretch ratios are now those of the $9:10$ coincidence cell. 

The difference from the $1:1$ case lies in the DSC lattice. 
For the $9:10$ interface, the shortest DSC translations are not pure in-plane dislocations with zero step height. 
Instead, the shortest admissible modes are disconnections with Burgers vectors in the interface plane. 
The relative translation also shifts the plane of coincidence sites, producing a nonzero step height. 
In the coordinate system defined above, the disconnection-mode matrix is 
\begin{equation}\label{eq:B_AuPd_9_10}
\underline{\mathbf{B}}
=
\left(\begin{array}{cccc}
\underline{\bm{b}}^{(1)} &
\underline{\bm{b}}^{(2)} &
\underline{\bm{b}}^{(3)} &
\underline{\bm{b}}^{(4)} 
\end{array}\right)
=
\left(\begin{array}{cccc}
\frac{\sqrt{10}}{20}\bar{a}_\text{Au} & \frac{\sqrt{10}}{20}\bar{a}_\text{Au} & 0 & 0 \\
0 & 0 & \frac{\sqrt{10}}{20}\bar{a}_\text{Au} & \frac{\sqrt{10}}{20}\bar{a}_\text{Au} \\
0 & 0 & 0 & 0 \\
\frac{1}{2}\bar{a}_\text{Au}' & -\frac{1}{2}\bar{a}_\text{Au}' & \frac{1}{2}\bar{a}_\text{Au}' & -\frac{1}{2}\bar{a}_\text{Au}'
\end{array}\right)
=
\frac{\sqrt{2} \bar{l}}{3}
\left(\begin{array}{cccc}
\frac{\sqrt{10}}{20} & \frac{\sqrt{10}}{20} & 0 & 0 \\
0 & 0 & \frac{\sqrt{10}}{20} & \frac{\sqrt{10}}{20} \\
0 & 0 & 0 & 0 \\
\frac{\kappa}{2} & -\frac{\kappa}{2}& \frac{\kappa}{2} & -\frac{\kappa}{2}
\end{array}\right),
\end{equation}
where $\bar{a}_\text{Au}$ and $\bar{a}_\text{Au}'$ are the lattice parameters of Au in the RS:
$\bar{a}_\text{Au} = (\sqrt{2}/3)\bar{l}$ and $\bar{a}_\text{Au}' = \bar{l}'$.
It is convenient to define the dimensionless ratio: $\kappa \equiv \bar{a}_\text{Au}' / \bar{a}_\text{Au} = (3/\sqrt{2})(\bar{l}'/\bar{l})$.  

Solving the qGFBE with Eq.~\eqref{uTpFijhnRiIhnothncssc} and Eq.~\eqref{eq:B_AuPd_9_10} gives the unique solution:
\begin{equation}
\mathbf{N}
=
\frac{3\sqrt{5}}{\bar{l}}
\left(\begin{array}{cccc}
c & c & -s & -s \\
s & s & c & c \\
0 & 0 & 0 & 0
\end{array}\right).
\end{equation}
Modes (1) and (2) form two arrays with the same density and the same line direction, while modes (3) and (4) form another pair of arrays orthogonal to the first pair. 
Because modes within each pair have opposite step heights, an equal-density combination produces no net step accumulation. 
Nevertheless, the elementary defects are disconnections rather than pure dislocations.

For mode (1), the angle between Burgers vector and the line direction is
\begin{align}
\omega^{(1)}
&=
\arccos\left(
\hat{\bm{b}}^{(1)} \cdot 
(\hat{\bm{\eta}}^{(1)} \times \hat{\bm{n}}_\txR)
\right)
=
\arccos\left(
\frac{s}{\sqrt{c^2 + s^2}}
\right)
=
\arccos\left(
\frac{2\sin(\Delta\theta/2)}
{\sqrt{(r_\text{Au} - r_\text{Pd})^2 \cos^2(\Delta\theta/2) + 4\sin^2(\Delta\theta/2)}}
\right)
\nonumber\\
&=
\frac{\pi}{2} - \frac{\Delta\theta}{2}
- \arccos\left(
\frac{(3/\sqrt{2})a_\text{Au} \sin(\Delta\theta)}{\sqrt{5}a_\text{Pd} - (3/\sqrt{2})a_\text{Au} \sin(\Delta\theta)}
\right).
\end{align}
The disconnection line spacing is
\begin{equation}\label{eq:d_AuPd_9_10}
d^{(1)}
=
\frac{1}{\left|\bm{\eta}^{(1)}\right|}
=
\frac{\bar{l} r_\text{Au} r_\text{Pd}}
{3\sqrt{5} \sqrt{r_\text{Au}^2 + r_\text{Pd}^2 - 2r_\text{Au} r_\text{Pd} \cos(\Delta\theta)}}
=
\frac{a_\text{Au} a_\text{Pd}}
{\sqrt{9a_\text{Au}^2 + 10a_\text{Pd}^2 - 6\sqrt{10}a_\text{Au} a_\text{Pd} \cos(\Delta\theta)}}.
\end{equation}
An interesting feature of Eq.~\eqref{eq:d_AuPd_9_10} is its behavior near $\Delta\theta=0$. Differentiating with respect to $\Delta\theta$ gives $(\rmd d(\Delta\theta) / \rmd \Delta\theta)_{\Delta\theta = 0} = 0$. 
Thus, the spacing varies smoothly around the special $9:10$ orientation relationship. 
This behavior is qualitatively different from the dislocation spacing in a low-angle twist or tilt GB. 
For example, the classical Frank formula for an array of GB dislocations gives $d = b/(2 \sin(\Delta\theta/2))$, which diverges as $\Delta\theta \to 0$. 
In that case, the reference orientation contains no lattice-parameter mismatch, so the required defect density vanishes when the misorientation vanishes. 
Consequently, the spacing has a cusp-like singularity at $\Delta\theta=0$.

\section{The interfaces involved in martensite transformation}

In this section, we apply the qGFBE to several interfaces associated with the martensite transformation between the high-temperature $\upbeta$ phase and the low-temperature $\upalpha$ phase.
The examples considered here are motivated by the crystallography of $\upbeta \to \upalpha$ transformations in titanium alloys, where the parent $\upbeta$ phase has a BCC structure and the product $\upalpha$ phase has an HCP structure.

The key distinction between the interfaces considered in this section and the GB or twist-interface examples discussed previously is that the two phases have different crystal structures. 
Therefore, even after choosing an orientation relationship, the two lattices cannot in general be brought into full three-dimensional coincidence by a rigid-body rotation alone. 
Instead, the construction of the RS necessarily involves homogeneous strains. 
The NS is then obtained by applying the inverse transformation to the strained lattice. The qGFBE determines whether the resulting interfacial incompatibility can be accommodated by a specified set of admissible disconnection modes.

We consider three representative interfacial geometries: the habit plane of a martensite nucleus, the broad face of a martensite nucleus, and the side face of a martensite nucleus. 
The first case illustrates how a single disconnection mode can accommodate the interface if the product phase orientation and interface inclination are allowed to adjust. 
The second case extends the analysis to the full three-dimensional transformation strain and shows that the solution forms a continuous set in the space of macroscopic interface geometries. 
The third case treats the side face and derives the condition under which a three-mode disconnection description is possible.

\subsection{Habit plane of a martensite nucleus}
\label{sec:onesethcpbcc}

\begin{figure}[tb]
\includegraphics[width=0.91\columnwidth]{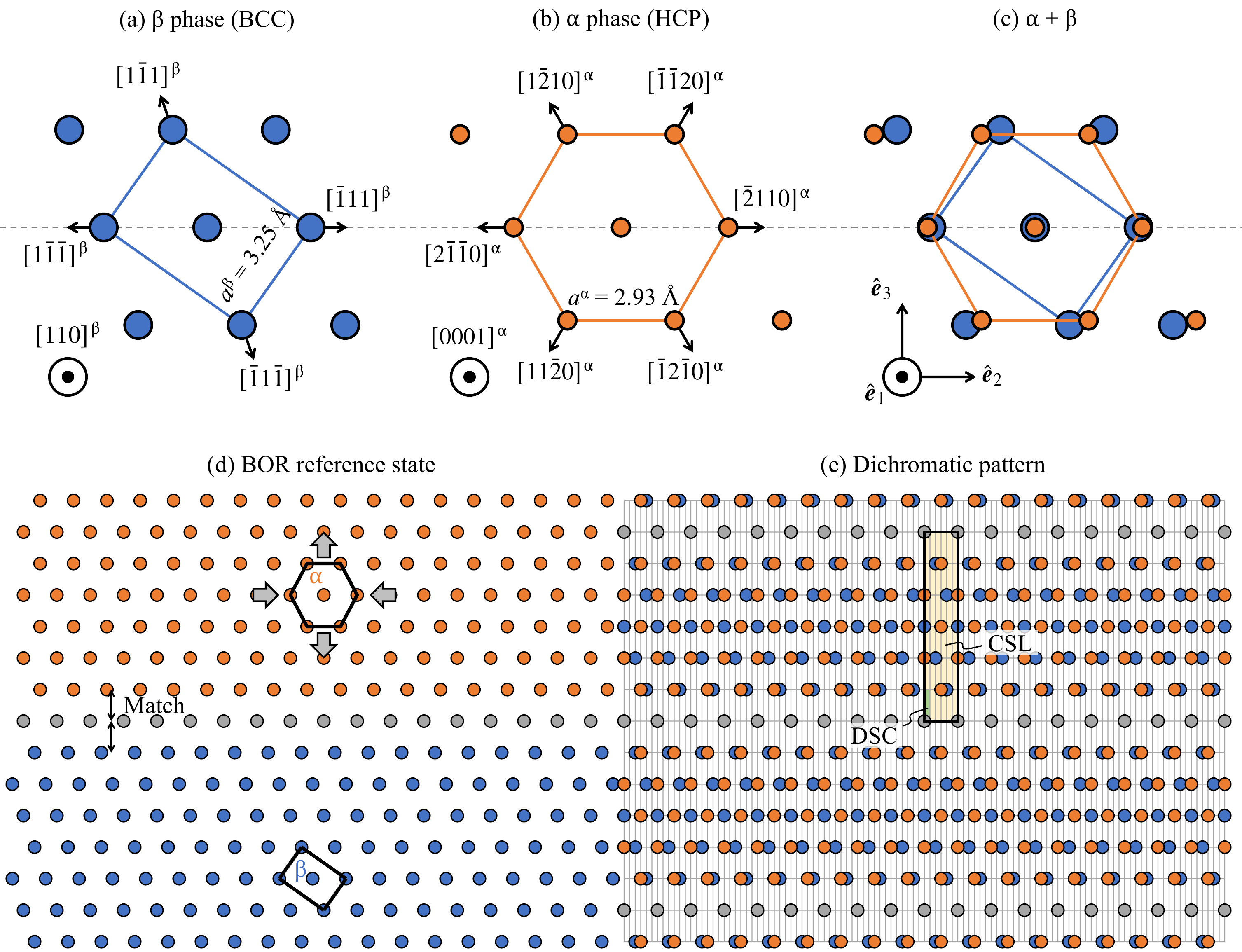}
\caption{\label{fig:alpha_beta_BOR}
Schematic of the Burgers orientation relationship between the $\upbeta$ (BCC) phase and the $\upalpha$ (HCP) phase. 
(a) Projection of the $(110)^\upbeta$ plane of the $\upbeta$ phase. 
(b) Projection of the $(0001)^\upalpha$ plane of the $\upalpha$ phase. 
(c) Superposition of (a) and (b), with $[110]^\upbeta$ aligned with $[0001]^\upalpha$ and  $[\bar{1}11]^\upbeta$ aligned with $[\bar{2}110]^\upalpha$.
The coordinate system is shown. 
In (a-c), gray dashed lines indicate the position of the interface.  
(d) Reference state obtained by straining the $\upalpha$ lattice (in orange) to match the $\upbeta$ lattice (in blue) along the interface plane (in gray). 
(e) Dichromatic pattern obtained by extending both lattices to fill the whole space. 
Gray points are coincidence sites. 
Gray grid denotes the DSC lattice. 
Yellow shaded region indicates a unit cell of the CSL. 
Green shaded region indicates a unit cell of the DSC lattice. 
} 
\end{figure}

We first consider the habit plane of a martensite nucleus near the Burgers orientation relationship (BOR) between the $\upbeta$ BCC phase and the $\upalpha$ HCP phase. 
The BOR is characterized by: $(110)^\upbeta \parallel (0002)^\upalpha$ and $[\bar{1}11]^\upbeta \parallel [\bar{2}110]^\upalpha$ (see Figs.~\ref{fig:alpha_beta_BOR}a-c).
The RS is constructed as follows. 
We take the $\upbeta$ phase as the reference lattice and homogeneously strain the $\upalpha$ phase so that it matches the $\upbeta$ phase along the interface plane $(1\bar{1}2)^\upbeta \parallel (0\bar{1}10)^\upalpha$.
The interplanar spacing normal to this interface is also matched in the RS. 
Thus, in the RS, the two lattices are coherent along the prescribed interface plane, as illustrated in Fig.~\ref{fig:alpha_beta_BOR}d.
We introduce the coordinate system: $\hat{\bm{e}}_1 \parallel [110]^\upbeta$, $\hat{\bm{e}}_2 \parallel [\bar{1}10]^\upbeta$, $\hat{\bm{e}}_3 \parallel [110]^\upbeta$. 

We take $\upbeta$ phase as the reference, and strain $\upalpha$ phase to match the $\upbeta$ phase along the interface plane $(1\bar{1}2)^\upbeta \parallel (0\bar{1}10)^\upalpha$ and also match the interplanar spacing; this is the RS (see Fig.~\ref{fig:alpha_beta_BOR}). 
Within the interface plane, the aligned directions are: $[110]^\upbeta \parallel [0001]^\upalpha$ and $[\bar{1}11]^\upbeta \parallel [\bar{2}110]^\upalpha$. 
Here, $\hat{\bm{n}}_\txR$ is the unit normal of the interface in the RS and is chosen to point from the $\upbeta$ phase to the $\upalpha$ phase.
We follow the conventional treatment of the martensite habit plane and neglect the lattice mismatch along the $\hat{\bm{e}}_1$ direction. 
This assumption imposes an invariant direction along $[110]^\upbeta \parallel [0001]^\upalpha$, which is necessary for the existence of an invariant plane. 
Since the $\upbeta$ phase is used as the reference, its deformation gradient is simply $\mathbf{F}^\upbeta = \mathbf{I}$. 
The stretch tensor that maps the $\upalpha$ phase from the RS back to its NS, or equivalently the inverse of the stretch used to strain the $\upalpha$ phase into registry with the $\upbeta$ phase in the RS, is
\begin{equation}\label{Ua1O2O3O2aaabO3aaab}
\mathbf{U}^\upalpha
=
\left(\begin{array}{ccc}
1 & 0 & 0 \\
0 & \Omega_2 & 0 \\
0 & 0 & \Omega_3 
\end{array}\right),
\quad
\Omega_2 
\equiv
\frac{2 a^\upalpha}{\sqrt{3} a^\upbeta}, 
\quad
\Omega_3
\equiv
\frac{3 a^\upalpha}{2\sqrt{2} a^\upbeta}, 
\end{equation}
where $a^\upalpha$ and $a^\upbeta$ are the lattice parameters of the $\upalpha$ and $\upbeta$ phases, respectively. 
The first diagonal component is set to unity because the mismatch along $\hat{\bm{e}}_1$ has been neglected.
The jump in the inverse deformation gradient is then $\llbracket \mathbf{F}^{-1} \rrbracket = (\mathbf{U}^\upalpha)^{-1} - \mathbf{I}$. 
If the interface inclination is assumed to remain unchanged from the RS to the NS, then $\hat{\bm{n}} = \hat{\bm{n}}_\text{R} = \hat{\bm{e}}_3$. 
Therefore, the tangential interfacial incompatibility tensor entering the qGFBE is
\begin{equation}\label{Tparallel_fixed_habit}
\underline{\mathbf{T}}^\parallel
=
\left(\begin{array}{c}
\llbracket \mathbf{F}^{-1} \rrbracket \\
\hat{\bm{n}}_\txR^T
\end{array}\right)
(\mathbf{I} - \hat{\bm{n}} \otimes \hat{\bm{n}})
=
\left(\begin{array}{ccc}
0 & 0 & 0 \\
0 & \Omega_2^{-1}-1 & 0 \\
0 & 0 & 0 \\
0 & 0 & 0 
\end{array}\right).
\end{equation}
From the CSL/DSC construction, the shortest Burgers-step vector relevant to this interface is $\underline{\bm{b}} = (0, (\sqrt{3}/12)a^\upbeta, 0, \sqrt{2/3}a^\upbeta)^T$. 
If we attempt to accommodate the interfacial mismatch using only this single disconnection mode, the disconnection-mode matrix is
\begin{equation}\label{uB0b0hab0s3}
\underline{\mathbf{B}}
=
\left(\begin{array}{c}
\underline{\bm{b}}
\end{array}\right)
=
\left(\begin{array}{c}
0 \\ b \\ 0 \\ h
\end{array}\right)
=
a^\upbeta
\left(\begin{array}{c}
0 \\
\sqrt{3}/12 \\
0 \\
\sqrt{2/3}
\end{array}\right).
\end{equation}
However, Eq.~\eqref{Tparallel_fixed_habit} is not column-wise parallel to $\underline{\mathbf{B}}$, because the nonzero column of $\underline{\mathbf{T}}^\parallel$ has no step component whereas the admissible disconnection mode does. Equivalently, $\text{rank}([\begin{array}{c|c} \underline{\mathbf{B}} & \underline{\mathbf{T}} \end{array}]) > \text{rank}(\underline{\mathbf{B}})$.
Therefore, the qGFBE has no solution for a fixed interface inclination if only the shortest single disconnection mode in Eq.~\eqref{uB0b0hab0s3} is allowed. 

\begin{figure}[tb]
\includegraphics[width=0.53\columnwidth]{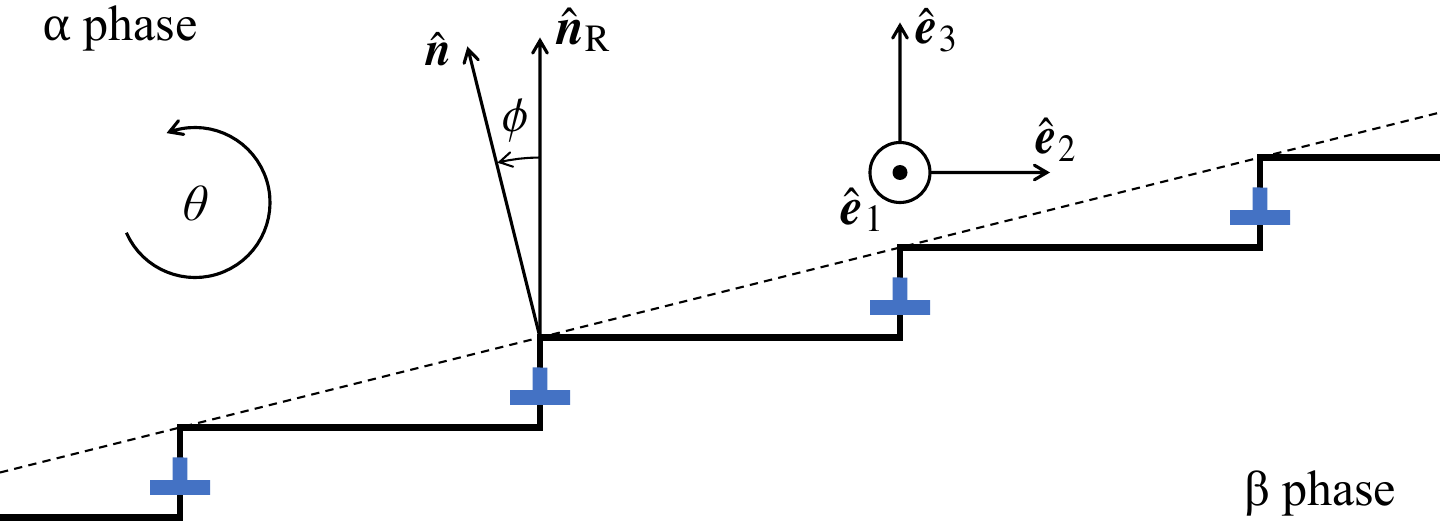}
\caption{\label{fig:alpha_beta_habit}
Schematic of the habit plane of the $\upalpha$/$\upbeta$ interface. 
The coordinate system refers to Fig.~\ref{fig:alpha_beta_BOR}c. 
The $\upalpha$ phase is allowed to rotate about the $\hat{\bm{e}}_1$ axis by an angle $\theta$, while the interface normal $\hat{\bm{n}}$ is allowed to rotate about the same axis by an angle $\phi$. 
Blue symbols denote disconnections. 
} 
\end{figure}

To obtain a solution, there are two possible strategies. 
One may either add more disconnection modes, thereby increasing the dimension of $\text{Col}(\underline{\mathbf{B}})$, or allow the macroscopic transformation geometry to vary, thereby changing $\underline{\mathbf{T}}^\parallel$. 
Since the goal here is to describe a habit plane accommodated by a single set of disconnections, we adopt the second strategy.
Specifically, we allow both the $\upalpha$ phase and the interface plane to rotate about the invariant direction $\hat{\bm{e}}_1$, as shown schematically in Fig.~\ref{fig:alpha_beta_habit}. 
The additional rotation of the $\upalpha$ phase is described by
\begin{equation}
\mathbf{R}^\upalpha
=
\left(\begin{array}{ccc}
1 & 0 & 0 \\
0 & \cos\theta & \sin\theta \\
0 & -\sin\theta & \cos\theta
\end{array}\right),
\end{equation}
where $\theta$ is the deviation of the $\upalpha$ orientation from the reference BOR. 
The rotated interface normal is written as
\begin{equation}
\hat{\bm{n}}
=
\left(\begin{array}{c}
0 \\ -\sin\phi \\ \cos\phi
\end{array}\right),
\end{equation}
where $\phi$ is the inclination angle of the habit plane relative to the original normal $\hat{\bm{n}}_\text{R}=\hat{\bm{e}}_3$. 
Thus, two additional unknowns, $\theta$ and $\phi$, are introduced.
With these additional degrees of freedom, the inverse deformation-gradient jump becomes $\llbracket \mathbf{F}^{-1} \rrbracket = (\mathrm{R}^\upalpha \mathbf{U}^\upalpha)^{-1} - \mathbf{I}$.
The interfacial incompatibility tensor is therefore
\begin{equation}
\underline{\mathbf{T}}^\parallel
=
\left(\begin{array}{c}
(\mathbf{R}^\upalpha \mathbf{U}^\upalpha)^{-1} - \mathbf{I}
\\
\hat{\bm{n}}_\txR^T
\end{array}\right)
(\mathbf{I} - \hat{\bm{n}} \otimes \hat{\bm{n}})
\nonumber\\
=
\left(\begin{array}{ccc}
\mathbf{0} & \underline{\bm{t}}_2^\parallel & \underline{\bm{t}}_3^\parallel
\end{array}\right),
\end{equation}
where
\begin{equation}\label{t3_habit_single_mode}
\underline{\bm{t}}_2^\parallel
=
\left(\begin{array}{c}
0 \\
(\Omega_2^{-1}\cos\theta - 1)\cos\phi - \Omega_2^{-1}\sin\theta \sin\phi \\
\Omega_3^{-1}\sin\theta \cos\phi + (\Omega_3^{-1}\cos\theta - 1)\sin\phi \\
\sin\phi 
\end{array}\right)
\cos\phi,
\quad
\underline{\bm{t}}_3^\parallel
=
\underline{\bm{t}}_2^\parallel \tan{\phi}.
\end{equation}
Equation~\eqref{t3_habit_single_mode} shows that the two nonzero columns of $\underline{\mathbf{T}}^\parallel$ are parallel. 
Consequently, the condition $\text{Col}(\underline{\mathbf{T}}^\parallel) \subseteq \text{Col}(\underline{\mathbf{B}})$ reduces to the simpler requirement: $\underline{\bm{t}}_2^\parallel \parallel \underline{\bm{b}}$.
Since $\underline{\bm{b}}$ has nonzero components only in the second and fourth entries, this parallelism condition requires that the third component of $\underline{\bm{t}}_2^\parallel$ vanish and that the ratio of the second and fourth components equal $b/h$. 
Hence, $\theta$ and $\phi$ must satisfy
\begin{equation}\label{habit_single_mode_nonlinear}
\left\{\renewcommand{\arraystretch}{2}
\begin{array}{l}
\Omega_3^{-1}\sin\theta \cos\phi + (\Omega_3^{-1}\cos\theta - 1)\sin\phi = 0
\\
\dfrac{(\Omega_2^{-1}\cos\theta - 1)\cos\phi - \Omega_2^{-1}\sin\theta \sin\phi}{\sin\phi}
=
\dfrac{b}{h}
\end{array}\right..
\end{equation}
This system of nonlinear equations determines the additional misorientation angle $\theta$ and the habit-plane inclination angle $\phi$ for a given lattice-parameter ratio.
Once $\theta$ and $\phi$ are obtained, the disconnection density follows directly from the qGFBE. Since only one disconnection mode is used, the density tensor contains a single vector, $\mathbf{N} = \bm{\eta}$.
The qGFBE requires $\underline{\mathbf{B}}\bm{\eta}^{T} = \underline{\mathbf{T}}^\parallel$.
Using the fourth component of $\bm{t}_2^\parallel$ in Eq.~\eqref{t3_habit_single_mode}, the magnitude of the disconnection density is $|\bm{\eta}| = |\sin\phi\cos\phi|/h$.
Therefore, the spacing between neighboring disconnection lines is
\begin{equation}
d
=
\frac{1}{|\bm{\eta}|}
=
\frac{h}{|\sin\phi\cos\phi|}.
\end{equation}

\begin{figure}[tb]
\includegraphics[width=0.83\columnwidth]{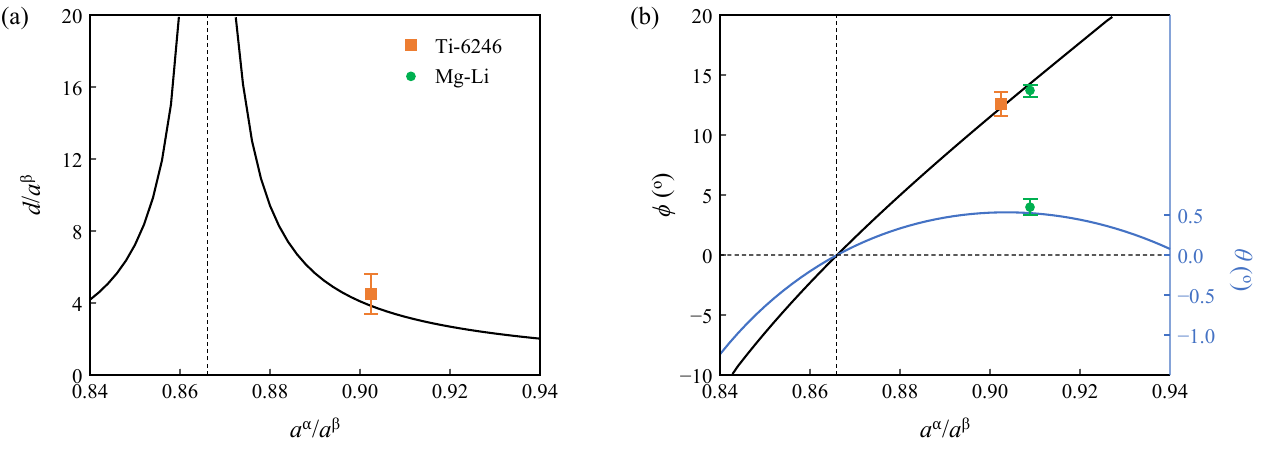}
\caption{\label{fig:abOneset_ratio}
Solution of the qGFBE for $\upalpha$/$\upbeta$ interfaces near the Burgers orientation relationship when the interfacial mismatch is accommodated by one set of disconnections. 
Orange squares and green circles are experimental results~\cite{Ackerman2020PRM, Kral2007Crystallography}. 
(a) Reduced disconnection line spacing vs. the lattice-parameter ratio. 
(b) Inclination angle $\phi$ (black curve) and misorientation angle $\theta$ (blue curve) vs. the lattice-parameters ratio.
} 
\end{figure}

The lattice parameters of the $\upalpha$ and $\upbeta$ phases depend on temperature and alloy composition. 
Therefore, the ratio $a^\upalpha/a^\upbeta$ varies from alloy to alloy and may also change during thermal processing. 
Figure~\ref{fig:abOneset_ratio} shows the solution of Eq.~\eqref{habit_single_mode_nonlinear} as a function of this lattice-parameter ratio. 
The calculated quantities include the reduced disconnection spacing $d/a^\upbeta$, the habit-plane inclination angle $\phi$, and the additional misorientation angle $\theta$. 
Experimental measurements from Refs.~\cite{Ackerman2020PRM,Kral2007Crystallography} are included for comparison and are consistent with the qGFBE prediction.

\subsection{Broad face of a martensite nucleus}
\label{sec:broadface}

The analysis in the previous subsection considered a simplified habit-plane construction in which the lattice mismatch along the direction $[110]^\upbeta \parallel [0001]^\upalpha$ direction was ignored. 
Equivalently, the first principal stretch in Eq.~\eqref{Ua1O2O3O2aaabO3aaab} was set to unity, $U^\upalpha_{11} = 1$.
This simplification is useful for deriving a one-disconnection-mode habit plane, but it does not account for the full transformation strain between the $\upbeta$ and $\upalpha$ phases. 
In this section, we retain the lattice mismatch in all three principal directions and analyze the broad face of a martensite nucleus.

The stretch tensor that maps the $\upalpha$ phase from the RS to the NS, or equivalently the inverse of the stretch applied to the $\upalpha$ phase to match the $\upbeta$ phase in the RS, is written as
\begin{equation}
\mathbf{U}^\upalpha
=
\left(\begin{array}{ccc}
\Omega_1 & 0 & 0 \\
0 & \Omega_2 & 0 \\
0 & 0 & \Omega_3 
\end{array}\right),
\quad
\Omega_1
\equiv
\frac{c^\upalpha}{\sqrt{2} a^\upbeta},
\quad
\Omega_2 
\equiv
\frac{2 a^\upalpha}{\sqrt{3} a^\upbeta}, 
\quad
\Omega_3
\equiv
\frac{3 a^\upalpha}{2\sqrt{2} a^\upbeta}, 
\end{equation}
where $a^\upbeta$ is the lattice parameter of the $\upbeta$ phase, while $a^\upalpha$ and $c^\upalpha$ are the lattice parameters of the $\upalpha$ phase. 
The three stretch ratios quantify the mismatch between the two phases along the three basis directions.
In addition to the full stretch, we now allow the orientation of the $\upalpha$ phase to vary in three dimensions. 
The rotation of the $\upalpha$ phase is represented by an axis-angle pair $(\hat{\bm{o}}, \theta)$￼, where $\hat{\bm{o}}$ is a unit rotation axis and $\theta$ is the rotation angle. 
Equivalently, the rotation tensor \(\mathbf{R}^{\upalpha}\) belongs to $\mathrm{SO}(3)$. Using Rodrigues' formula,
\begin{equation}
\mathbf{R}^\upalpha
=
\cos\theta \mathbf{I}
+
(1-\cos\theta) (\hat{\bm{o}} \otimes \hat{\bm{o}})
+
\sin\theta (\hat{\bm{o}} \times),
\end{equation}
with $\hat{\bm{o}} \cdot \hat{\bm{o}} = 1$.
The interface inclination is also allowed to vary freely. 
We describe the interface normal in the NS by $\hat{\bm{n}} = (\hat{n}_1, \hat{n}_2, \hat{n}_3)^T$ with $\hat{\bm{n}} \cdot \hat{\bm{n}} = 1$. 
The interfacial incompatibility tensor entering the qGFBE is then
\begin{equation}\label{uTpUainvRaTI001Ihnothn}
\underline{\mathbf{T}}^\parallel
=
\left(\begin{array}{c}
(\mathbf{U}^\upalpha)^{-1} (\mathbf{R}^\upalpha)^T - \mathbf{I} \\
(0, 0, 1)
\end{array}\right)
(\mathbf{I} - \hat{\bm{n}} \otimes \hat{\bm{n}}),
\end{equation}
which contains five degrees of freedom.

In the one-mode habit-plane analysis, only the shortest Burgers-step vector was included. 
Once the full three-dimensional lattice mismatch is retained, one disconnection mode is generally insufficient. 
We therefore include one additional disconnection mode. 
The disconnection-mode matrix becomes
\begin{equation}
\underline{\mathbf{B}}
=
\left(\begin{array}{cc}
\underline{\bm{b}}^{(1)} & \underline{\bm{b}}^{(2)}
\end{array}\right)
=
a^\upbeta
\left(\begin{array}{cc}
0 & \frac{\sqrt{2}}{2} \\
\frac{\sqrt{3}}{12} & \frac{\sqrt{3}}{6} \\
0 & \frac{\sqrt{6}}{6} \\
\frac{\sqrt{6}}{3} & 0
\end{array}\right).
\end{equation}

The qGFBE has a solution if and only if every column of $\underline{\mathbf{T}}^\parallel$ lies in the column space of $\underline{\mathbf{B}}$: $\text{Col}(\underline{\mathbf{T}}^\parallel) \subseteq \text{Col}(\underline{\mathbf{B}})$.
Because $\underline{\bm{b}}^{(1)}$ and $\underline{\bm{b}}^{(2)}$ are linearly independent vectors in $\mathbb{R}^4$, they span a two-dimensional plane in the four-dimensional Burgers-step space.
Hence, the above column-space condition can be checked by using the left nullspace of $\underline{\mathbf{B}}$. 
Any vector in this left nullspace is orthogonal to both Burgers-step vectors and therefore orthogonal to the entire column space of $\underline{\mathbf{B}}$. 
Let $\underline{\bm{b}}^* \in \mathbb{R}^4$ be such a left-null vector. 
Then,  
\begin{equation}
(\underline{\bm{b}}^*)^T \underline{\mathbf{B}} = \mathbf{0}
\quad \Leftrightarrow \quad
\underline{\bm{b}}^* \cdot \underline{\bm{b}}^{(1)} = 0
\text{ and }
\underline{\bm{b}}^* \cdot \underline{\bm{b}}^{(2)} = 0.
\end{equation} 
In practice, the left nullspace can be found by a QR decomposition of $\underline{\mathbf{B}}$:
\begin{equation}
\underline{\mathbf{B}}_{4\times 2} 
=
\mathbf{Q}_{4\times 4} 
\left(\begin{array}{c}
\mathbf{R}_{2\times 2} \\
\mathbf{0}_{2 \times 2}
\end{array}\right)
=
\left(\begin{array}{cccc}
\bm{q}_1 & \bm{q}_2 & \bm{q}_3 & \bm{q}_4
\end{array}\right)
\left(\begin{array}{cc}
* & * \\
0 & * \\
0 & 0 \\
0 & 0
\end{array}\right).
\end{equation}
Here, $\mathbf{Q}_{4\times 4}$ is an orthogonal tensor with all columns orthonormal; and $\mathbf{R}_{2\times 2}$ is an upper triangular matrix. 
Immediately, we find that $\text{span}(\bm{q}_1, \bm{q}_2) = \text{span}(\underline{\bm{b}}^{(1)}, \underline{\bm{b}}^{(2)}) = \text{Col}(\underline{\mathbf{B}})$. 
Because $\bm{q}_3$ and $\bm{q}_4$ are orthogonal to $\bm{q}_1$ and $\bm{q}_2$ and they are unit vectors, $\bm{q}_3$ and $\bm{q}_4$ are the basis vectors of the left nullspace of $\underline{\mathbf{B}}$, i.e., $\text{span}(\bm{q}_3, \bm{q}_4) = \text{Null}(\underline{\mathbf{B}}^T)$. 
Thus, $\underline{\bm{b}}^{*(1)}$ and $\underline{\bm{b}}^{*(2)}$ (not necessarily orthonormal) are two linear combinations of $\bm{q}_3$ and $\bm{q}_4$. 
A convenient choice is
\begin{equation}
\underline{\bm{b}}^{*(1)}
=
\left(\begin{array}{c}
1 \\ 0 \\ -\sqrt{3} \\ 0
\end{array}\right)
\text{ and }
\underline{\bm{b}}^{*(2)}
=
\left(\begin{array}{c}
0 \\ -4\sqrt{2} \\ 4 \\ 1
\end{array}\right).
\end{equation}
If the qGFBE has a solution, then all columns of $\underline{\mathbf{T}}^\parallel$ must be linear combinations of $\underline{\bm{b}}^{(1)}$ and $\underline{\bm{b}}^{(2)}$ (and thus linear combinations of $\bm{q}_1$ and $\bm{q}_2$).
Therefore, every column of $\underline{\mathbf{T}}^\parallel$ must be orthogonal to both left-null vectors. 
This gives 
\begin{equation}\label{ubsTuTpzuTp}
(\underline{\bm{b}}^{*(m)})^T \underline{\mathbf{T}}^\parallel = \mathbf{0}
\quad\Rightarrow\quad
(\underline{\mathbf{T}}^\parallel)^T \underline{\bm{b}}^{*(m)} 
=
(\mathbf{I} - \hat{\bm{n}} \otimes \hat{\bm{n}}) \underline{\mathbf{T}}^T \underline{\bm{b}}^{*(m)}
= \mathbf{0}
\quad
\text{for $m=1,2$.}
\end{equation}
Define $\bm{v}^{(m)} \equiv \underline{\mathbf{T}}^T \underline{\bm{b}}^{*(m)}$; obviously, $\bm{v}^{(m)} = \bm{v}^{(m)}(\hat{\bm{o}}, \theta)$.
Equation~\eqref{ubsTuTpzuTp} requires that the projection of $\bm{v}^{(m)}$ onto the interface plane should be zero, i.e., $\bm{v}^{(m)} \parallel \hat{\bm{n}}$. 
Equivalently, there exist two scalar multipliers $\lambda^{(1)}$ and $\lambda^{(2)}$ such that
\begin{equation}
\bm{v}^{(1)}(\hat{\bm{o}}, \theta)
= \lambda^{(1)} \hat{\bm{n}}
\text{ and }
\bm{v}^{(2)}(\hat{\bm{o}}, \theta)
= \lambda^{(2)} \hat{\bm{n}}.
\end{equation}
These two vector equations represent six scalar nonlinear equations. 
The two multipliers, $\lambda^{(1)}$ and $\lambda^{(2)}$, are auxiliary unknowns introduced only to express the parallelism conditions.
Explicitly, 
\begin{equation}
\bm{v}^{(1)}(\hat{\bm{o}}, \theta)
=
\left(\renewcommand{\arraystretch}{1.5}
\begin{array}{c}
\Omega_3^{-1} \left[(1-\cos\theta)\hat{o}_1 \hat{o}_3 + \sin\theta \hat{o}_2\right]
-
\frac{\sqrt{3}}{3} \left[\Omega_1^{-1} (\cos\theta + (1-\cos\theta)\hat{o}_1^2) - 1\right]
\\
\Omega_3^{-1} \left[(1-\cos\theta)\hat{o}_2 \hat{o}_3 - \sin\theta \hat{o}_1\right]
-
\frac{\sqrt{3}}{3} \Omega_1^{-1} \left[(1-\cos\theta)\hat{o}_1 \hat{o}_2 + \sin\theta \hat{o}_3\right]
\\
\Omega_3^{-1} \left[\cos\theta + (1-\cos\theta) \hat{o}_3^2\right] - 1
-
\frac{\sqrt{3}}{3} \Omega_1^{-1} \left[(1-\cos\theta) \hat{o}_1 \hat{o}_3 - \sin\theta \hat{o}_2\right]
\end{array}\right),
\end{equation}
\begin{equation}
\bm{v}^{(2)}(\hat{\bm{o}}, \theta)
=
\left(\renewcommand{\arraystretch}{1.5}
\begin{array}{c}
\frac{4\sqrt{3}}{3} \left[\Omega_1^{-1} (\cos\theta + (1-\cos\theta)\hat{o}_1^2) - 1\right] 
-
4\sqrt{2} \Omega_2^{-1} \left[(1-\cos\theta)\hat{o}_1\hat{o}_2 - \sin\theta \hat{o}_3\right]
\\
\frac{4\sqrt{3}}{3} \Omega_1^{-1} \left[(1-\cos\theta)\hat{o}_1 \hat{o}_2 + \sin\theta \hat{o}_3\right] 
-
4\sqrt{2} \left[\Omega_2^{-1}(\cos\theta + (1-\cos\theta)\hat{o}_2^2) - 1\right]
\\
\frac{4\sqrt{3}}{3} \Omega_1^{-1} \left[(1-\cos\theta)\hat{o}_1 \hat{o}_3 - \sin\theta \hat{o}_2\right] 
-
4\sqrt{2} \Omega_2^{-1} \left[(1-\cos\theta)\hat{o}_2\hat{o}_3 + \sin\theta \hat{o}_1\right] + 1
\end{array}\right).
\end{equation}

The condition for the existence of a qGFBE solution can be summarized as
\begin{equation}
\left\{\begin{array}{l}
\bm{v}^{(1)}(\hat{\bm{o}}, \theta)
= \lambda^{(1)} \hat{\bm{n}}
\\
\bm{v}^{(2)}(\hat{\bm{o}}, \theta)
= \lambda^{(2)} \hat{\bm{n}}
\\
\hat{\bm{o}} \cdot \hat{\bm{o}} = 1
\\
\hat{\bm{n}} \cdot \hat{\bm{n}} = 1
\end{array}\right..
\end{equation}
This is a system of eight scalar equations for nine scalar unknowns: $\hat{o}_1$, $\hat{o}_2$, $\hat{o}_3$, $\theta$, $\hat{n}_1$, $\hat{n}_2$, $\hat{n}_3$, $\lambda^{(1)}$, and $\lambda^{(2)}$.
Equivalently, if the rotation is regarded intrinsically as an element of $\mathrm{SO}(3)$, the interface normal is regarded as an element of $\mathrm{S}^2$, and the multipliers $\lambda^{(1)}$ and $\lambda^{(2)}$ are eliminated, then the solution is described by four scalar equations in a five-dimensional macroscopic geometry space. 
One possible form of these equations is
\begin{equation} 
\left\{\begin{array}{l}
v^{(1)}_1(\hat{\bm{o}}, \theta) \hat{n}_3 - v^{(1)}_3(\hat{\bm{o}}, \theta) \hat{n}_1 = 0
\\
v^{(1)}_2(\hat{\bm{o}}, \theta) \hat{n}_3 - v^{(1)}_3(\hat{\bm{o}}, \theta) \hat{n}_2 = 0
\\
v^{(2)}_1(\hat{\bm{o}}, \theta) \hat{n}_3 - v^{(2)}_3(\hat{\bm{o}}, \theta) \hat{n}_1 = 0
\\
v^{(2)}_2(\hat{\bm{o}}, \theta) \hat{n}_3 - v^{(2)}_3(\hat{\bm{o}}, \theta) \hat{n}_2 = 0
\end{array}\right.
\quad
\text{for }
(\hat{\bm{o}}, \theta) \in \mathrm{SO}(3), ~
\hat{\bm{n}} \in \mathrm{S}^2.
\end{equation}
The qGFBE solution for the broad face forms a one-dimensional solution set, or curve, in the five-dimensional space $(\hat{\bm{o}}, \theta, \hat{\bm{n}}) \in \mathrm{SO}(3)\times\mathrm{S}^2$. 
This result shows that, once the full lattice mismatch is included, the broad face of a martensite nucleus is characterized not by a unique habit-plane geometry, but by a continuous family of compatible interface geometries. 
Additional energetic criteria are therefore needed to select the particular interface geometry observed in experiment.

\subsection{Side face of a martensite nucleus}
\label{sec:sideface}

\begin{figure}[tb]
\includegraphics[width=0.91\columnwidth]{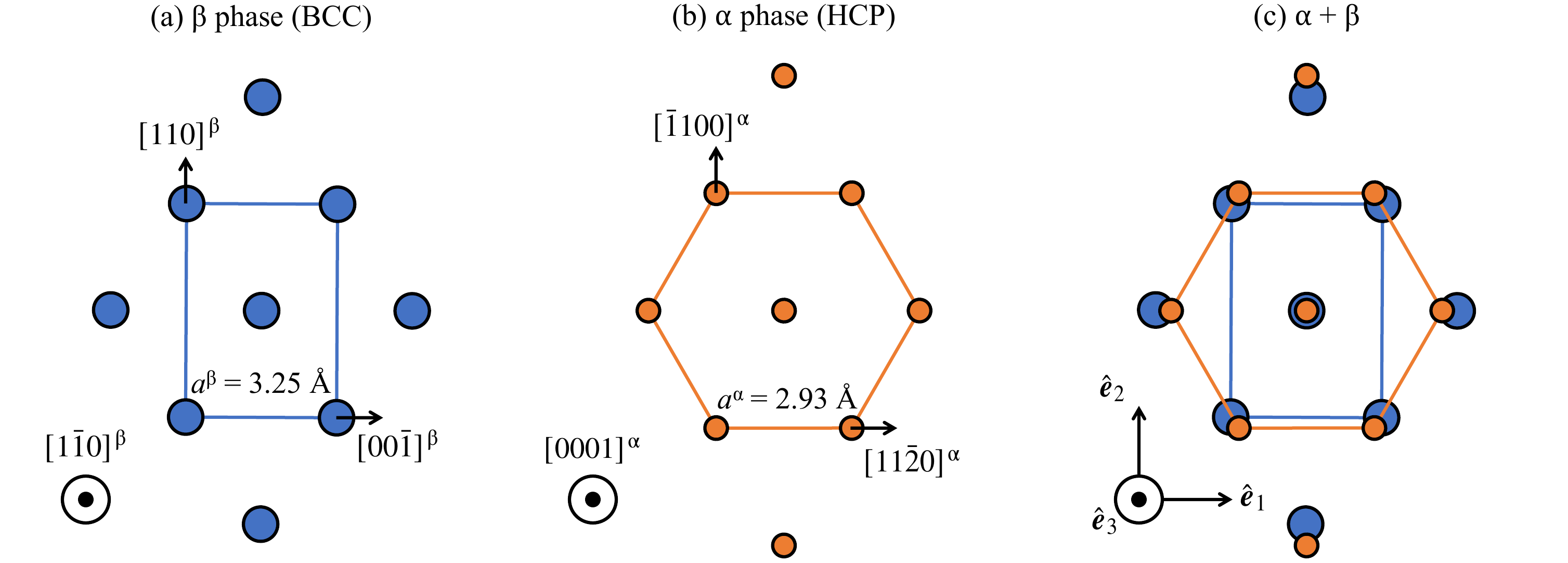}
\caption{\label{fig:alpha_beta_PSOR}
Schematic of Pitsch-Schrader orientation relationship between $\upbeta$ (BCC) and $\upalpha$ (HCP) phases. 
(a) Projection of $(1\bar{1}0)^\upbeta$ plane of $\upbeta$ phase. 
(b) Projection of $(0001)^\upalpha$ plane of $\upalpha$ phase. 
(c) Superimposition of (a) and (b) with $[00\bar{1}]^\upbeta$ aligned with $[11\bar{2}0]^\upalpha$ and  $[110]^\upbeta$ aligned with $[\bar{1}100]^\upalpha$.
The coordinate system is established. 
} 
\end{figure}

The side face of the martensite nucleus is analyzed using the Pitsch-Schrader orientation relationship (PSOR) between the parent $\upbeta$ (BCC) phase and the product $\upalpha$ (HCP) phase. 
A schematic of this orientation relationship is shown in Fig.~\ref{fig:alpha_beta_PSOR}.
The PSOR reference state is constructed by aligning the following crystallographic directions: $[00\bar{1}]^\upbeta \parallel [11\bar{2}0]^\upalpha$, $[110]^\upbeta \parallel [\bar{1}100]^\upalpha$, and $[1\bar{1}0]^\upbeta \parallel [0001]^\upalpha$.
The $\upalpha$ lattice is then homogeneously strained to match the $\upbeta$ lattice in this reference configuration. 
The interface plane is chosen to have normal $\hat{\bm{n}}_\txR \parallel [1\bar{1}0]^\upbeta$, which is the direction normal to the projected plane shown in Fig.~\ref{fig:alpha_beta_PSOR}. 
This completes the construction of the RS.
Based on the RS, we introduce the coordinate system: $\hat{\bm{e}}_1 \parallel [00\bar{1}]^\upbeta$, $\hat{\bm{e}}_2 \parallel [110]^\upbeta$, $\hat{\bm{e}}_3 \parallel [1\bar{1}0]^\upbeta$, as shown in Fig.~\ref{fig:alpha_beta_PSOR}c.

The stretch tensor that maps the $\upalpha$ phase from its RS to NS is
\begin{equation}\label{Ua1O2O3O2aaabO3aaab}
\mathbf{U}^\upalpha
=
\left(\begin{array}{ccc}
\Omega_1 & 0 & 0 \\
0 & \Omega_2 & 0 \\
0 & 0 & \Omega_3 
\end{array}\right),
\quad
\Omega_1
\equiv
\frac{a^\upalpha}{a^\upbeta}, 
\quad
\Omega_2 
\equiv
\frac{\sqrt{3} a^\upalpha}{\sqrt{2} a^\upbeta}, 
\quad
\Omega_3
\equiv
\frac{c^\upalpha}{\sqrt{2} a^\upbeta}.
\end{equation}
As in the broad-face analysis, both the orientation of the $\upalpha$ phase and the interface inclination are allowed to vary. 
The orientation change of the $\upalpha$ phase is represented by a rotation $\mathbf{R}^\upalpha \in \mathrm{SO}(3)$, and the interface normal in the NS is represented by $\hat{\bm{n}} \in \mathrm{S}^2$.
The interfacial incompatibility tensor has the same form as that used for the broad face; see Eq.~\eqref{uTpUainvRaTI001Ihnothn}. 

Based on the CSL/DSC lattice of the PSOR reference state, the relevant DSC lattice vectors are equivalent to lattice vectors of the $\upbeta$ phase. 
The disconnection modes considered here are collected in the matrix
\begin{equation}
\underline{\mathbf{B}}
=
a^\upbeta
\left(\begin{array}{ccc}
\frac{1}{2} & 1 & 0 \\
-\frac{\sqrt{2}}{2} & 0 & 0 \\
0 & 0 & 0 \\
0 & 0 & \sqrt{2}
\end{array}\right).
\end{equation}
The first two modes are pure dislocation modes with in-plane Burgers vectors and zero step height, whereas the third mode is a pure step mode.
The column space of $\underline{\mathbf{B}}$ is $\text{Col}(\underline{\mathbf{B}}) = \text{span}(\hat{\bm{e}}_1, \hat{\bm{e}}_2, \hat{\bm{e}}_4)$; so, $\text{rank}(\underline{\mathbf{B}}) = 3$. 
The qGFBE has a solution if and only if the interfacial incompatibility tensor lies in the column space of the admissible disconnection modes, namely, $\text{Col}(\underline{\mathbf{T}}^\parallel) \subseteq \text{Col}(\underline{\mathbf{B}})$. 
Equivalently, $\text{rank}([\begin{array}{c|c} \underline{\mathbf{B}} & \underline{\mathbf{T}}^\parallel \end{array}]) = 3$.
Because the third row of $\underline{\mathbf{B}}$ is identically $\mathbf{0}$, the third row of $\underline{\mathbf{T}}^\parallel$ must vanish if the above column-space condition is to be satisfied.
The third row of $\underline{\mathbf{T}}^\parallel$ is $\bm{t}^\parallel_\text{r3} = \hat{\bm{e}}_3^T \llbracket \mathbf{F}^{-1} \rrbracket (\mathbf{I} - \hat{\bm{n}} \otimes \hat{\bm{n}})$. 
\begin{equation}
\hat{\bm{e}}_3^T \llbracket \mathbf{F}^{-1} \rrbracket 
=
\hat{\bm{e}}_3^T 
\left[(\mathbf{U}^\upalpha)^{-1} (\mathbf{R}^\upalpha)^T - \mathbf{I}\right]
=
\hat{\bm{e}}_3^T 
\left(\Omega_3^{-1} {\mathbf{R}^\upalpha}^T - \mathbf{I}\right)
\quad\Rightarrow\quad
(\bm{t}^\parallel_\text{r3})^T
=
(\mathbf{I} - \hat{\bm{n}} \otimes \hat{\bm{n}})
\left(\Omega_3^{-1} \mathbf{R}^\upalpha - \mathbf{I}\right) \hat{\bm{e}}_3,
\end{equation}
If solution exists, or equivalently, $\text{rank}([\begin{array}{c|c} \underline{\mathbf{B}} & \underline{\mathbf{T}}^\parallel \end{array}]) = 3$,
\begin{equation}\label{kpr3zIhnothnOeinvRaIhe3}
\bm{t}^\parallel_\text{r3}
= \mathbf{0}
\quad\Rightarrow\quad
(\mathbf{I} - \hat{\bm{n}} \otimes \hat{\bm{n}})
\left(\Omega_3^{-1} \mathbf{R}^\upalpha - \mathbf{I}\right) \hat{\bm{e}}_3
=
\bm{0}
\quad\Rightarrow\quad
\left(\Omega_3^{-1} \mathbf{R}^\upalpha - \mathbf{I}\right) \hat{\bm{n}}_\txR
\parallel \hat{\bm{n}}. 
\end{equation}
Therefore, 
\begin{equation}
\text{rank}([\begin{array}{c|c} \underline{\mathbf{B}} & \underline{\mathbf{T}}^\parallel \end{array}])
=
\left\{\begin{array}{ll}
3, & \left(\Omega_3^{-1} \mathbf{R}^\upalpha - \mathbf{I}\right) \hat{\bm{n}}_\txR
\parallel \hat{\bm{n}}
\\
4, & \text{otherwise}
\end{array}\right..
\end{equation}
Thus, the qGFBE has a solution only when $\left(\Omega_3^{-1} \mathbf{R}^\upalpha - \mathbf{I}\right) \hat{\bm{n}}_\txR
\parallel \hat{\bm{n}}$.
When this condition is satisfied, the solution is unique for a prescribed macroscopic geometry $(\mathbf{R}^{\upalpha},\hat{\bm{n}})$. 
The solution is 
\begin{align}\label{NuTpTuBTse1e2e3}
&\mathbf{N} 
= 
(\underline{\mathbf{T}}^\parallel)^T (\underline{\mathbf{B}}^T)^*
~\Rightarrow~
\left\{\begin{array}{l}
\bm{\eta}^{(1)}
=
-\frac{\sqrt{2}}{a^\upbeta} 
(\mathbf{I} - \hat{\bm{n}} \otimes \hat{\bm{n}})
\left(\Omega_2^{-1} \mathbf{R}^\upalpha - \mathbf{I}\right) \hat{\bm{e}}_2
\\
\bm{\eta}^{(2)}
=
\frac{1}{a^\upbeta} 
(\mathbf{I} - \hat{\bm{n}} \otimes \hat{\bm{n}})
\left[
\left(\Omega_1^{-1} \mathbf{R}^\upalpha - \mathbf{I}\right) \hat{\bm{e}}_1
+
\frac{\sqrt{2}}{2} 
\left(\Omega_2^{-1} \mathbf{R}^\upalpha - \mathbf{I}\right) \hat{\bm{e}}_2
\right]
\\
\bm{\eta}^{(3)}
=
\frac{\sqrt{2}}{2a^\upbeta}
(\mathbf{I} - \hat{\bm{n}} \otimes \hat{\bm{n}})
\hat{\bm{e}}_3
\end{array}\right..
\end{align}
These expressions must always be interpreted together with the condition $\left(\Omega_3^{-1} \mathbf{R}^\upalpha - \mathbf{I}\right) \hat{\bm{n}}_\txR
\parallel \hat{\bm{n}}$.
Without this condition, the tangential incompatibility tensor has a component outside the column space of the selected disconnection modes, and the qGFBE has no solution.

The degree-of-freedom count is as follows. 
The three $\{\bm{\eta}^{(m)}\}$ vectors contain nine scalar components. 
The rotation $\mathbf{R}^{\upalpha}$ contributes three independent degrees of freedom, and the interface normal ￼$\hat{\bm{n}}$ contributes two. 
Thus, the total number of unknowns is $9+3+2 = 14$.
The qGFBE (Eq.~\eqref{NuTpTuBTse1e2e3}) provides the nine scalar equations. 
The compatibility condition provides two additional independent scalar constraints (because it requires one vector to be parallel to another in three-dimensional space). 
Hence, the total number of independent equations is $9+2=11$.
The problem is therefore underdetermined at this stage.

Experimental observations indicate that the three disconnection arrays on the side face are approximately parallel. 
This condition can be written as $\bm{\eta}^{(1)} \parallel \bm{\eta}^{(2)} \parallel \bm{\eta}^{(3)}$. 
Since all three vectors lie in the same interface plane with normal $\hat{\bm{n}}$￼, requiring them to be mutually parallel supplies two additional independent scalar constraints. 
Including these constraints gives $11+2 = 13$ independent equations for $14$ unknowns. 
Therefore, the solution is expected to form a one-dimensional curve in the high-dimensional space of admissible side-face geometries.

\bibliography{GFBE}